\documentclass[12pt]{article}
\usepackage{array}
\usepackage{tabularx}
\usepackage{amsmath,amssymb,amsfonts,epsfig,cite,setspace,bigstrut,framed,eufrak}
\usepackage[all]{xy}
\usepackage{color}
\usepackage{pifont}
\usepackage{xcolor}

\usepackage{mathtools}

\makeatletter \@addtoreset{equation}{section} \makeatother
\renewcommand{\theequation}{\thesection.\arabic{equation}}

\begin{document}

\vskip 0.25in

\newcommand{\todo}[1]{{\bf\color{blue} !! #1 !!}\marginpar{\color{blue}$\Longleftarrow$}}
\newcommand{\nn}{\nonumber}
\newcommand{\comment}[1]{}
\newcommand\T{\rule{0pt}{2.6ex}}
\newcommand\B{\rule[-1.2ex]{0pt}{0pt}}

\newcommand{\CO}{{\cal O}}
\newcommand{\cI}{{\cal I}}
\newcommand{\cM}{{\cal M}}
\newcommand{\cW}{{\cal W}}
\newcommand{\cN}{{\cal N}}
\newcommand{\cR}{{\cal R}}
\newcommand{\cH}{{\cal H}}
\newcommand{\cK}{{\cal K}}
\newcommand{\cT}{{\cal T}}
\newcommand{\cZ}{{\cal Z}}
\newcommand{\cO}{{\cal O}}
\newcommand{\cQ}{{\cal Q}}
\newcommand{\cB}{{\cal B}}
\newcommand{\cC}{{\cal C}}
\newcommand{\cD}{{\cal D}}
\newcommand{\cE}{{\cal E}}
\newcommand{\cF}{{\cal F}}
\newcommand{\cA}{{\cal A}}
\newcommand{\cX}{{\cal X}}
\newcommand{\IA}{\mathbb{A}}
\newcommand{\IP}{\mathbb{P}}
\newcommand{\IQ}{\mathbb{Q}}
\newcommand{\IH}{\mathbb{H}}
\newcommand{\IR}{\mathbb{R}}
\newcommand{\IC}{\mathbb{C}}
\newcommand{\IF}{\mathbb{F}}
\newcommand{\IS}{\mathbb{S}}
\newcommand{\IV}{\mathbb{V}}
\newcommand{\II}{\mathbb{I}}
\newcommand{\IZ}{\mathbb{Z}}
\newcommand{\re}{{\rm Re}}
\newcommand{\im}{{\rm Im}}
\newcommand{\tr}{\mathop{\rm Tr}}
\newcommand{\ch}{{\rm ch}}
\newcommand{\rk}{{\rm rk}}
\newcommand{\ext}{{\rm Ext}}
\newcommand{\bi}{\begin{itemize}}
\newcommand{\ei}{\end{itemize}}
\newcommand{\beq}{\begin{equation}}
\newcommand{\eeq}{\end{equation}}
\newcommand{\bea}{\begin{eqnarray}}
\newcommand{\eea}{\end{eqnarray}}
\newcommand{\ba}{\begin{array}}
\newcommand{\ea}{\end{array}}

\newcommand{\CN}{{\cal N}}
\newcommand{\y}{{\mathbf y}}
\newcommand{\z}{{\mathbf z}}
\newcommand{\C}{\mathbb C}\newcommand{\R}{\mathbb R}
\newcommand{\CA}{\mathbb A}
\newcommand{\CP}{\mathbb P}
\newcommand{\cP}{\mathcal P}
\newcommand{\tmat}[1]{{\tiny \left(\begin{matrix} #1 \end{matrix}\right)}}
\newcommand{\mat}[1]{\left(\begin{matrix} #1 \end{matrix}\right)}
\newcommand{\diff}[2]{\frac{\partial #1}{\partial #2}}
\newcommand{\gen}[1]{\langle #1 \rangle}

\newtheorem{theorem}{\bf THEOREM}
\newtheorem{proposition}{\bf PROPOSITION}
\newtheorem{observation}{\bf OBSERVATION}

\def\theequation{\thesection.\arabic{equation}}
\newcommand{\setall}{
	\setcounter{equation}{0}
}
\renewcommand{\thefootnote}{\fnsymbol{footnote}}

\begin{titlepage}
\vfill
\begin{flushright}
{\tt\normalsize KIAS-P22061}\\

\end{flushright}
\vfill
\begin{center}
{\Large\bf Holonomy Saddles and 5d BPS Quivers}

\vskip 1.5cm

Qiang Jia and Piljin Yi
\vskip 5mm

{\it School of Physics,
Korea Institute for Advanced Study, Seoul 02455, Korea}

\end{center}
\vfill

\begin{abstract}

We study the Seberg-Witten geometry of 5d $\cN=1$ pure Yang-Mills
theories compactified on a circle. The concept of the holonomy saddle
implies that there are multiple 4d limits of interacting Seiberg-Witten
theories from a single 5d theory, and we explore this in the simplest
case of pure $SU(N)$ theories. The compactification leads to $N$
copies of locally indistinguishable 4d pure $SU(N)$ Seiberg-Witten
theories in the infrared, glued together in a manner dictated
by the Chern-Simons level. We show how this picture naturally builds
the 5d BPS quivers which agree with the D0 probe dynamics previously
proposed via the geometrically engineered local Calabi-Yau.
We work out various $SU(2)$ and $SU(3)$ examples through a detailed
look at the respective spectral curves. We also note a special $\IZ_{2N}$
feature of $SU(N)_N$ spectral curves and the resulting BPS quivers,
with emphasis on how the 4d holonomy saddles are affected.

\end{abstract}

\vfill
\end{titlepage}

\tableofcontents

\section{Overview}

In studying gauge theories compactified on a circle, the holonomy
along the latter
\bea
Pe^{ i \int_{\IS^1}A}
\eea
often plays a central role in understanding
the infrared nature of the theory. Take, for instance, the confinement
and de-confinement phase transition of pure Yang-Mills theory in the
finite temperature. The observable that discerns the confining phase
is the vacuum expectation value
\bea
\langle {\rm tr}\,Pe^{ i \int_0^\beta A}\rangle
\eea
along the thermal circle; If the theory confines, this observable,
written in the defining representation of $SU(N)$ has to vanish since
a stand-alone quark cannot exist as a finite energy state. The transition between the two phases is expected as we change the temperature across a critical value.

We may attribute such behaviors to the temperature-dependent quantum effective potential
for the holonomy variable. In the confining phase, the lowest energy configuration
of the potential must be such that the eigenvalues of $A$ along the circle are distributed democratically along the holonomy circle so that its exponentiated values cancel one another.
If the vacuum is not confining, on the other hand, the eigenvalues prefer to be typically
clustered around a single position, so that the exponentiated value would add up rather
than canceling one another.

Although the holonomy represents classically flat directions,
the perturbative and non-perturbative quantum effects would generally lift such degeneracy
and favor particular distribution of eigenvalues. This phenomenon of the holonomy
variables acquiring quantum effective potential persists with 4d $\cN=1$ supersymmetry,
where pure Yang-Mills theory is expected to be confining again in low enough temperature.
Demonstrating this scenario analytically is one of the most coveted goals in theoretical
physics.

On the other hand, one can also ask a simpler question of what happens if the circular
direction is not taken as a thermal circle but considered as a supersymmetric compactification. Operationally, one would achieve this by demanding the periodic boundary condition on fermions
instead of the anti-periodic one associated with the thermal circle. The answer is well-known \cite{Davies:1999uw};
the quantum effective superpotential emerges at a nonperturbative level \cite{Lee:1997vp}, and can be kept track
of quite rigorously. In fact, this remains the best known demonstration of the confinement
on the space-time $\IR^3\times\IS^1$.

Generally, one may encounter more than one preferred holonomy vacua. One class of 4d
object that shows this cleanly is the A-twisted partition function \cite{Closset:2017bse}
for models with non-anomalous $U(1)$ R-symmetry; the latter symmetry is used to twist
supercharge such that supersymmetric partition function can be written for a spacetime manifold which
is $T^2$ fibred over general Riemannian surface. In these classes of theories, the discrete
vacua for the $T^2$ holonomy demanded by the effective potential can be shown to be determined
by a certain modular function. When one of the two circles is taken to a small size, turning
the theory effectively to 3d $\cN=2$, one finds these discrete solutions cluster at several
values, each of which gives an effective 3d supersymmetric theory \cite{Hwang:2018riu}.

We refer to this kind of phenomenon where interacting lower dimensional theories emerge
at certain preferred holonomy values as the holonomy saddle \cite{Hwang:2017nop}.
One place where the holonomy saddle enters rather crucially is in the computation
of the Witten index \cite{Witten:1982df}, which, for 4d, is the same
as the above A-twisted index with the simple spacetime, $T^4$. The notion
of the holonomy saddle enters physics in how the Witten index of a $d$ dimensional theory
is expressible as a sum of several Witten indices of different $d-1$ dimensional theories.
One hallmark of such holonomy saddles in the context of the Witten index is the absence of
decoupled $U(1)$ factor in the zero radius limit in the reduced lower dimensional theory
\cite{Hwang:2017nop,Duan:2020qjy},
since the latter's free gaugino would kill the contribution to the bulk index.

With larger supersymmetries, on the other hand, the supersymmetric vacua can prove to be
continuously numerous. Take, for example, 4d $\cN=2$ Seiberg-Witten theory \cite{Seiberg:1994rs,Seiberg:1994aj} where
a continuous set of superselection sectors are labeled by the Coulombic vacuum expectation
values. The Witten index no longer makes sense and the holonomy saddles would take
a little different meaning. Consider our main objective, i.e., 5d $\cN=1$ theory
on a circle. If the holonomy along the circle takes a generic value and if the
circle size is infinitesimal, this would break the gauge symmetry to the Cartan at
a scale $1/R_5$, with all charged particles at such a mass scale, and one ends up
with a product of free Abelian theories as the effective 4d theories which are
rather uninteresting.

As such, we will extend the notion of the holonomy saddle for these higher supersymmetric
cases as those special corners of the holonomy torus where 5d Seiberg-Witten theory reduces to
non-Abelian and interacting 4d Seiberg-Witten theories. In general, the effective
4d Seiberg-Witten theory and its 5d origin need not share the same field content;
one can expect the former to form a subsector of the latter in general.
The simplest example would be 5d $\cN=1$ $SU(2)$ theories with fundamental flavors of
light masses $\sim m \ll 1/R_5$. At $R_5A_5\sim 0$, one finds 4d $SU(2)$ theory with
the same flavors, while at the opposite end $R_5A_5\sim \sigma_3/2$, one would find a pure
$SU(2)$ theory since there the fundamental quarks would have acquired a large mass
of order $ 1/R_5$ and decouple in the small radius limit \cite{Ganor:1996pc}.

In the context of the Witten index computation for gauged quantum mechanics,
it is known that pure Yang-Mills theory typically produces multiple holonomy saddles
\cite{Hwang:2017nop}. With gauge group $G$, distinct holonomy
saddles are classified by the maximal non-Abelian subgroups of $G$ \cite{Hwang:2017nop}.
The classification originates from the Witten index counting and has resolved
an old puzzle \cite{Staudacher:2000gx,Pestun:2002rr} on various D-brane bound state
counting in the presence of the orientifold \cite{Lee:2017lfw}. As noted already,
it is correlated to the absence of decoupled $U(1)$
in the small radius limit, so it remains relevant for our extended notion of the
holonomy saddles for 5d Seiberg-Witten theories on a circle.
For the general Lie group, the maximal non-Abelian subgroup may be constructed
by cutting a single node from the affine Dynkin diagram, so their variety
for a given $G$ goes like $r_G +1$ with $r_G={\rm rank}\;G$. As such, there
are at least $r_G+1$ 4d Seiberg-Witten theories that
can be found in the small radius limit of a single 5d pure $G$ Seiberg-Witten theory.

$SU(N)$, on the other hand,  has only one type of maximal non-Abelian subgroup, namely
$SU(N)$ itself. This does not mean a unique holonomy saddle, however. We would
find $N=r_{SU(N)}+1$
identical copies of such $SU(N)$ holonomy saddles, evenly separated and related by
the left multiplications by the center $\IZ_N$ of  $SU(N)$ along the holonomy
torus  \cite{Hwang:2017nop}. These copies would look identical to each other
if we ignore how they are glued together in the holonomy torus.
If the $SU(N)$ Chern-Simons level is turned off, this center symmetry $\IZ_N$ remains
intact, so there is another option of viewing the theory as $SU(N)/\IZ_N$ with
a single holonomy saddle. No local observable can detect a difference between
two such choices. If one chooses to keep all $N$ saddles as distinct, the
gauge volume becomes $N$-fold larger, so we effectively multiply by $N$
only to divide by $N$ again at the end.\footnote{One should take care not to confuse the discussion
here with what is commonly called ``$G/\Gamma$ theories" in recent literature.
The latter involves either 2-cycles in the spacetime along which Stiefel-Whitney
classes are invoked and the resulting topologically nontrivial bundles are summed over
in the path integral \cite{Witten:2000nv}, or certain external objects, which
cannot be built as a coherent object from the local fields, are inserted
in the path integral \cite{Aharony:2013hda}.}

With Chern-Simons level $k$, on the other hand, this latter
option shrinks, $\IZ_N\rightarrow \IZ_{{\rm gcd}(N,k)}$,
as we will see later more explicitly. Given these different global behaviors with different
$k$, it makes sense to associate the theory with the simply-connected $SU(N)$
to accommodate all these possibilities in a single framework, instead of
minimizing down to $SU(N)/\IZ_{{\rm gcd}(N,k)}$ case by case. This universal
viewpoint leads, for example, to the description where 5d $SU(N)_0$ is composed
of a $N$-fold covering, each copy of which contains a single 4d $SU(N)$ in the
small radius limit. In Ref.~\cite{Jia:2021ikh},
the authors took the opposite viewpoint
for $SU(2)_0$ where a division by $\IZ_2$ halved the holonomy torus. More generally,
the Chern-Simons level controls how these $N$ 4d saddles are glued together along
the holonomy torus to form the full 5d Seiberg-Witten geometry.


One motivation for studying the 5d Seiberg-Witten theory on a small circle is the
5d BPS quiver \cite{Closset:2019juk}.
This gauged quantum mechanics governs the BPS spectrum of such
compactified 5d theory and generalizes the more familiar 4d BPS quivers. This
constructive approach is especially suitable for the small radius limit since
here the magnetic and the electric BPS objects enter on equal footing; Recall
how from a 5d viewpoint the magnetic objects are string-like while electric ones
are point-like and only upon compactification on a circle the two types of
charges can be both realized via particle-like states. The 5d nature adds
two additional conserved charges, namely the Pontryagin charge and the KK charge,
so the 5d BPS quiver picks up two more nodes, relative to its 4d cousin.
The resulting 5d BPS quiver is in a sense more immediate than its 4d cousins,
from the viewpoint of geometrical engineering, as it happens to be the
single D0 probe theory of the relevant local Calabi-Yau. On the other hand, the question of
how such a 5d BPS quiver is derived from a pure field theory perspective
has been unavailable.

One objective of this note is by understanding how
multiple 4d Seiberg-Witten theories are embedded in a single 5d Seiberg-Witten
theory. Each 4d limit, i.e., each holonomy saddle, comes with its own 4d BPS
quivers, yet, in trying to smoothly interpolate one from another, we find
that the nodes of a pair of such 4d BPS quiver cannot be generally identified.
Rather, one finds some of the 4d BPS dyons in one saddle pick up a KK charge and
Pontryagin charges in the next holonomy saddle. This gives us a natural way
to add two mode nodes to a given 4d BPS quiver, as needed to complete the 5d BPS quiver.

Of course, the precise map and thus the precise shape of the 5d BPS quiver
depends on how one moves from one saddle to the next, and we identify the
path that constructs the canonical, via the toric diagram, D0 probe theory
this way. Iterate routes would build different shapes of the quiver, related
to the canonical one by quiver mutations.

This note is organized as follows. In Sec.2, we study the toric geometry
that builds the local Calabi-Yau for $SU(N)_k$. The spectral curve for
the Seiberg-Witten geometry, for circle-compactified theories, follows
immediately and we also note a subtlety in how we deal with the meromorphic
differential. Given these data, we dote on special corners of the holonomy
torus that generates 4d Seiberg-Witten theories in the small radius limit
and point out how $k$-dependence enters the relations among these holonomy
saddles.

In Sec.3, we specialize in $SU(2)$ theories and keep track of how various BPS states transform as we follow them
continuously from one saddle to the next. As is well known, the BPS spectra
of 4d $SU(2)$ Seiberg-Witten theory can be built up from a pair of states,
a monopole and a dyon in their simplest form. As we move from one 4d saddle
to the other, we will find that these basis states pick up the KK charge
and the instanton charge appropriately. Stated backward this means that
the usual pair of a monopole and a dyon in the other saddle would come
from BPS states in the first saddle that are equipped with the instanton
charge and the KK charge in some specific manner. These two extra BPS
states in the first saddle, together with the canonical 4d BPS states, span
a quiver diagram, which precisely reproduces the 5d BPS quiver previously
obtained from the D0 probe dynamics.

In this process, the discrete theta angle enters how the two 4d saddles
are glued together for the full 5d Seiberg-Witten geometry, which in
turn affects the 5d BPS quiver thus constructed, and results in the
well-known D0 probe dynamics for F0 and F1 geometry, respectively.
With the observation that the discrete theta angle of $SU(2)$ theory
is fundamentally the same thing as the Chern-Simons level of $SU(N>2)$
theories, this exercise lift to $SU(3)_k$ quite naturally. We devote
Sec.4 to the latter exercise and comment on the symmetry of the quiver diagram in Sec.5.  In the appendix, we outline a previous
related work in Ref.~\cite{Closset:2021lhd}, which in effect performed a monodromy
analysis that led to a mutated version of the 5d BPS quiver.

\section{5d Seiberg-Witten and Holonomy Saddles}

We will be dealing with $SU(2)$ and $SU(3)$ theories mostly in this note. There is one well-known
topological difference between these two classes in that $SU(3)$ theory admits non-Abelian
Chern-Simons level whereas $SU(2)$ instead allows discrete $\theta$ angle associated with
$\pi_4(SU(2))=\IZ_2$ \cite{Witten:1982fp}. The latter cannot be written as a local term in the effective action
but its effect manifests in the spectrum, such as how electric charges are conferred to the
instanton soliton.

However, these two types of topological couplings may be
considered on equal footing, as can be seen by realizing that both can be
generated by coupling a fundamental hyper and taking its mass $m$ to infinite.
Integrating out a massive hypermultiplet actually generates an eta invariant
\begin{equation}
	\Delta S_{\textrm{eff}} \sim \pi\,\eta(A),
\end{equation}
whose local part, i.e. the part that depends on $A$ continuously is the Chern-Simons
action \cite{Alvarez-Gaume:1984zst,Witten:2015aba}. The precise coefficient depends on how we take the Pauli-Villars regulator
field, but the rule of thumb is that each hypermultiplet in the fundamental
representation generates half of the above, which translates to half of the unit-quantized
Chern-Simons term as well. For $SU(2)=Sp(1)$ and in fact for $Sp(k)$ more generally,
this continuous part of the eta invariant is absent, leaving behind the global
and quantized part, rendering its would-be Chern-Simons coefficient to be
$\IZ_2$ valued; the same can be viewed as the discrete $\theta$ angle
for $Sp(k)$ theory. It is therefore not surprising that, for many purposes, the
roles played by the Chern-Simons level for $SU(N>2)$ would be emulated by the discrete
theta angle of $SU(2)$ and vice versa.

The main question we wish to address in this note is exactly how these topological
5d terms, with no analog in 4d, manifest as we compactify 5d theory on a small
circle. If we take the naive dimensional reduction of a 5d Seiberg-Witten theory
with the gauge group $G$, one should expect again a 4d $G$ Seiberg-Witten theory.
However, the point with the holonomy saddles was that there is in general more
than one possible 4d Seiberg-Witten theory which appears at the infrared end
if we take a more careful treatment and view entire the holonomy torus.

On the other hand, as we already mentioned in the first section, a unique
aspect of pure $SU(N)$ theories is that all holonomy saddles would carry the same
gauge group $SU(N)$, related by the center-shift and thus locally indistinguishable.
Indeed we will in this section see how this comes about by taking a close look at
the spectral curves and the resulting Seiberg-Witten geometries. We will find $N$
holonomy saddles which are locally identical to one another while these inherently
5d topologically couplings enter this picture in how these identical copies
of 4d $SU(N)$ theories would glue together to form the genuine 5d theory.
We will deal with these questions in more depth in later sections, a byproduct
of which is the 5d BPS quivers.

\subsection{The Spectral Curves from the Toric Data}

Let us briefly review the 5d $N=1$ supersymmetric gauge theory \cite{Seiberg:1996bd,Morrison:1996xf,Douglas:1996xp,Katz:1996fh,Intriligator:1997pq,
Aharony:1997bh,Jefferson:2017ahm,Jefferson:2018irk,Closset:2018bjz} and we refer to  \cite{Jia:2021ikh} for various conventions. The Coulomb phase of the theory is parametrized by the expectation value of a real scalar field $\phi$ in the vector multiplet. On a generic point in the Coulomb phase, the gauge group $G$ is broken to its Cartan part ${\rm U}(1)^{r_G}$, where $r_G$ is the rank of the group $G$. The dynamics on the Coulomb phase are encoded in the so-called Intriligator-Morrison-Seiberg (IMS) prepotential  \cite{Intriligator:1997pq}, which is one-loop exact.

It is convenient to geometrically engineer the 5d theory using M-theory compactified on a local Calabi-Yau 3-fold $\hat{X}$, which is a crepant resolution of the singular Calabi-Yau 3-fold $X$ \cite{Katz:1996fh,Intriligator:1997pq,Closset:2018bjz}. The resolved geometry contains various 4-cycles (also called divisors) and 2-cycles. The rank of the gauge group $r_G$ equals the number of the compact divisors $S_i (i=1,\cdots,r_G)$ and the M5-branes wrapped on $S_i$ give the 5d monopole strings charged under the {\rm U}(1)$_i$. On the other hand, M2-branes wrapped on the compact 2-cycle $C$ give BPS particles, and the electric charges under $\textrm{{\rm U}(1)}_i$ are given by the intersection numbers $C\cdot S_i$ inside the Calabi-Yau $\hat{X}$.

\begin{figure}[pbth]
\centering
\includegraphics[scale=0.35]{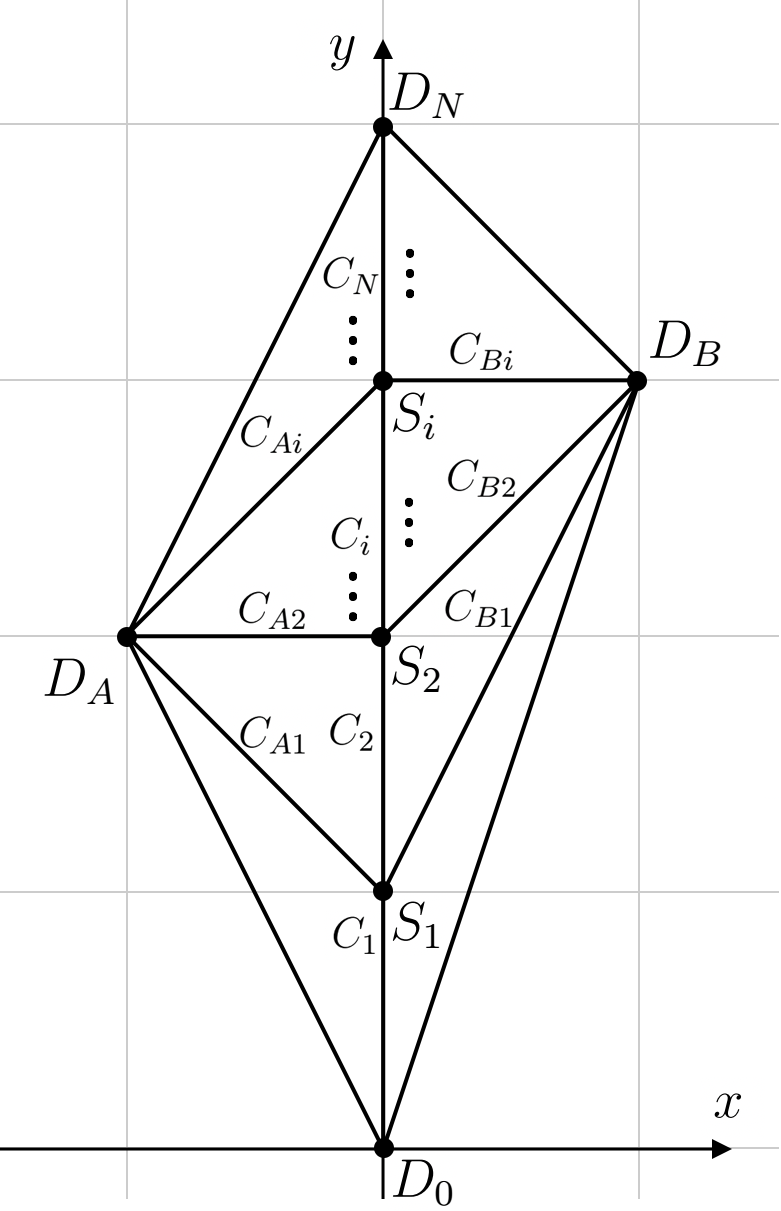}
\caption{\small The toric diagram for pure $SU(N)$ theory.}
\label{Fig-toric-grid}
\end{figure}

We will focus on the toric local Calabi-Yau 3-folds whose geometries are encoded in toric diagrams. For example, the toric Calabi-Yau 3-fold $\hat{X}$ that gives rise to 5d pure $SU(N)$ theories is shown in Fig.~\ref{Fig-toric-grid}, which is a subset of 2-dimensional lattice $\mathbb{Z}^2$. Each node represents a divisor in $\hat{X}$ and among them the external nodes $D_0,D_N,D_A,D_B$ correspond to non-compact divisors and the internal nodes $S_1,S_2,\cdots,S_{N-1}$ correspond to compact divisors. The internal lines $C_{Ai},C_{Bi}$ and $C_i$ are compact 2-cycles and they are the intersections of the two divisors associated with the two endpoints of the line.

After compactification on a circle with radius $R_5$ the prepotential will receive contributions from the membrane instantons and the theory becomes much more complicated \cite{Katz:1996fh,Lawrence:1997jr}. Also, with the Coulomb vev in the strongly coupled region, which would be below the KK-scale $1/2\pi R_5$, the membrane instanton sum is divergent and must be resumed. One way to probe such a strongly coupled region is to use mirror symmetry, mapping the type IIA string theory compactified on the Calabi-Yau 3-fold $\hat{X}$ to a type IIB string theory compactified on the mirror Calabi-Yau 3-fold $\hat{X}'$.

The mirror Calabi-Yau 3-fold \cite{Candelas:1990rm,Hori:2000kt} $\hat{X}'$ is given by a hypersurface in $\mathbb{C}^2 \times (\mathbb{C}^*)^2$ with the equation \cite{Closset:2021lhd}
\begin{equation}
	\hat{X}' = \{v_1 v_2 + P(t_1,t_2) = 0\ | \ (v_1,v_2)\in \mathbb{C},(t_1,t_2)\in (\mathbb{C}^*)^2 \},
\end{equation}
where $P(t_1,t_2)$ is a polynomial which can be read directly from the toric diagram
\begin{equation}\label{Eq-mirror-curve}
	P(t_1,t_2) = \sum_{m\in \Gamma_0} c_m t_1^{x_m} t_2^{y_m},
\end{equation}
and the sum is over all nodes $\Gamma_0 \subset \mathbb{Z}^2 $ in the toric diagram with coordinates $(x_m,y_m)$. The coefficients $c_m$ parametrize the complex structure of the mirror Calabi-Yau subject to the redundancy
\begin{equation}
	P(t_1,t_2) \sim s_0 P(s_1 t_1, s_2 t_2).\quad (s_0,s_1,s_2 \in \mathbf{C}^*)
\end{equation}
Those complex structure parameters of the mirror Calabi-Yau 3-fold $\hat{X}'$ are related to the $\ddot{\textrm{K}}$ahler structure parameters of $\hat{X}$ via the mirror map, as usual.

Given a curve $C$ in $\hat{X}$, the complex volume is given by
\begin{equation}
	t_C = \int_C (B + i J),
\end{equation}
here $B$ is the anti-symmetric 2-form in IIA theory and $J$ is the K$\ddot{\textrm{a}}$hler form of $\hat{X}$. On the other hand, the mirror parameter $z_C$ is given by
\begin{equation}
z_C = \prod_{m\in\Gamma_0} (c_m)^{C \cdot D_m},
\end{equation}
where $C\cdot D_m$ is the intersection number between 2-cycle $C$ with the divisor $D_m$ represented by each node.

The mirror map associates $t_C$ and $z_C$ such that in the asymptotic region of the K$\ddot{\textrm{a}}$hler moduli space (large volume limit) one has
\begin{equation}
t_C \approx \frac{1}{2\pi i} \log(z_C) + \mathcal{O}(z_C).
\end{equation}
The spectral curve $\Sigma$ is defined by the polynomial $P(t_1,t_2)$ as
\begin{equation}
	\Sigma = \{P(t_1,t_2)=0\} \in (\mathbb{C}^*)^2,
\end{equation}
and the periods of the holomorphic 3-form on $\hat{X}'$ can be reduced to an integral along the 1-cycles on the spectral curve $\Sigma$
\begin{equation}
	\Pi_{\gamma} = \int_{\gamma} \lambda_{\textrm{SW}},
\end{equation}
which gives the central charges of various BPS states, where
the Seiberg-Witten differential $\lambda_{\textrm{SW}}$ obeys
\begin{equation}\label{eq-Seiberg-Witten-differential}
	d \lambda_{\textrm{SW}} = \frac{1}{i(2\pi)^2R_5 } \frac{d t_1}{t_1} \wedge \frac{d t_2}{t_2}.
\end{equation}
As usual, $\lambda_{\textrm{SW}}$ is itself ambiguous and
here this ambiguity goes a little
beyond the one we are familiar with in the 4d setting.

\begin{figure}[phth]
\centering
\includegraphics[scale=0.25]{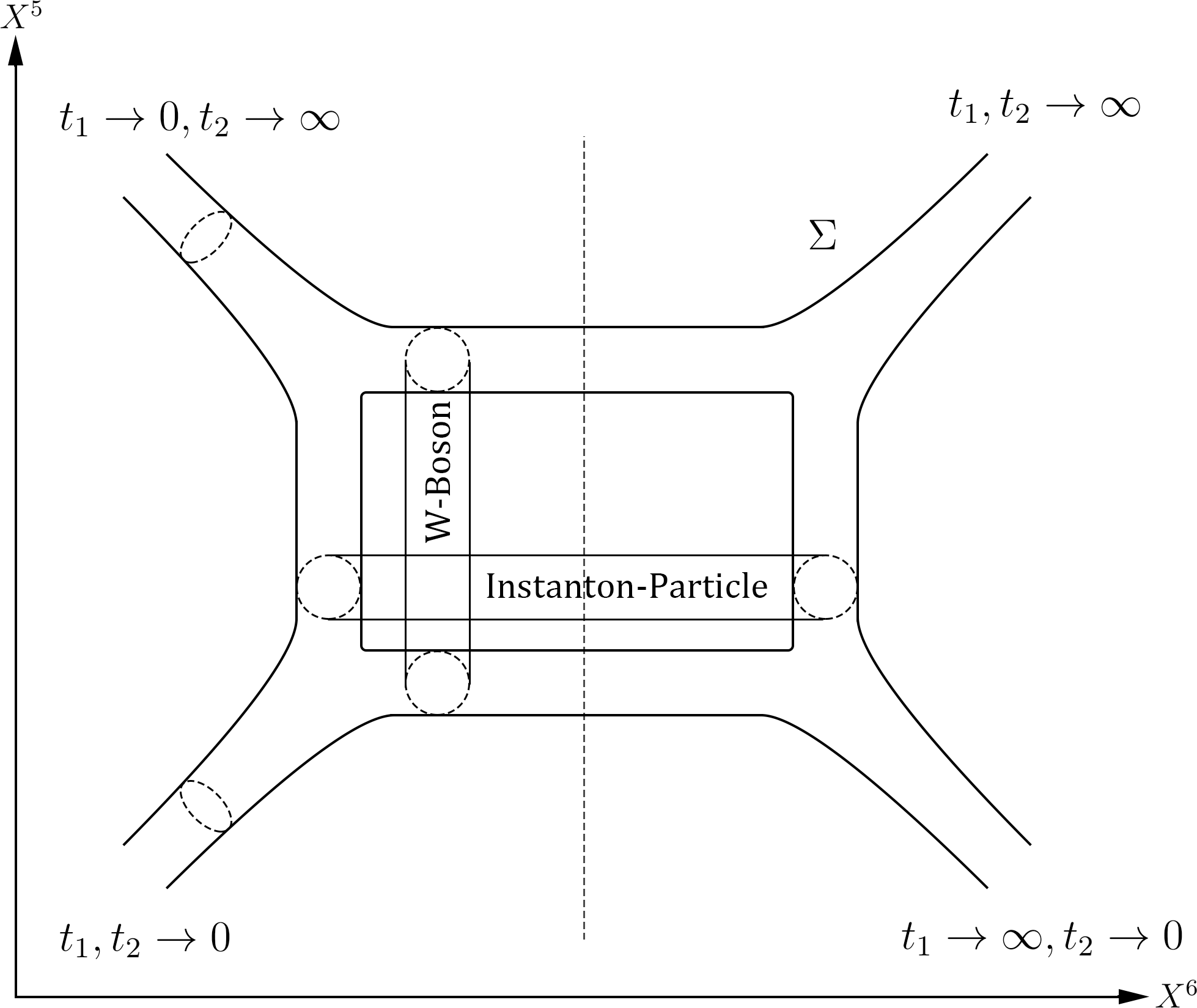}
\caption{\small The spectral curve $\Sigma$ for $SU(2)_0$ theory. The M5-brane is wrapping the spectral curve $\Sigma$. The W-boson and instanton particle are M2-branes that end on a pair of 1-cycles on the spectral curve as shown in the figure. In the 5d limit, the circular direction of the spectral curve will close and it becomes the brane web for $SU(2)_0$ theory. }
\label{Fig-5dSU20-brane-web}
\end{figure}

For this, it is useful to recall yet another construction, which
relies on M5-brane \cite{Witten:1997sc}. Pure Yang-Mills may be constructed as a
worldvolume dynamics on an M5-brane in flat spacetime spanned by
$X^{0,\cdots,9}$ and the M-theory direction $X^{11}$, compactified
on two circles, say, along $X^5$ and $X^{11}$ of radii $R_{\textrm{M},5}$
and $R_{\textrm{M},11}$ respectively. Parameterizing $t_1,t_2$ as
\begin{equation}
	t_1 = e^{\frac{X^6+i X^{11}}{R_{\textrm{M},11}}},\quad t_2 = e^{\frac{X^4+ i X^5}{R_{\textrm{M},5}}},
\end{equation}
a single M5-brane wrapped the above spectral curve $P(t_1,t_1)=0$
reproduces the Seiberg-Witten theory.\footnote{IIB $g_s$
and the compactification radius $R_5$  are related to this M5 brane side as,
$g_s = R_{\textrm{M},11}/R_{\textrm{M},5}$ and $R_5 = \alpha'/R_{\textrm{M},5}$, such that
\begin{equation}
t_1=e^{2\pi R_5 T_{D1}(X^6+i X^{11})},\quad t_2=e^{2\pi R_5 T_{F1}(X^4+i X^{5})},
\end{equation}
where $T_{D1}/T_{F1}$ is the tension of D/F-string. Also, note that
$R_5\sim l_P^3/(R_{\textrm{M},5}R_{\textrm{M},11})$
with the M-theory Planck length $l_P$. }
See Fig.~\ref{Fig-5dSU20-brane-web} for an illustration for $SU(2)_0$
theory. This realization can be connected to the above local Calabi-Yau
construction via IIB $(p,q)$ 5-brane web \cite{Aharony:1997bh} compactified on a circle,
as is well known.

The BPS states are associated with M2-branes that end on the M5-brane
and their central charges can be written as integrals of the
holomorphic 2-form pulled back onto the M2-brane \cite{Henningson:1997hy,Mikhailov:1997jv}
\begin{equation}
	Z_{\textrm{BPS}} =-iR_{\textrm{M},5}R_{\textrm{M},11}T_{\rm M2}
\int_{\textrm{M2}} \frac{d t_1}{t_1} \wedge \frac{d t_2}{t_2}
\end{equation}
with the integrand being part of the three K$\ddot{\textrm{a}}$hler forms
\bea
{\rm Re}\left(\frac{d t_1}{t_1} \wedge \frac{d t_2}{t_2}\right)\ ,\quad
{\rm Im}\left(\frac{d t_1}{t_1} \wedge \frac{d t_2}{t_2}\right)\ ,\quad
\frac{i}{2}\left(\frac{d t_1}{t_1} \wedge \frac{d \bar t_1}{\bar t_1}+\frac{d t_2}{t_2} \wedge \frac{d \bar t_2}{\bar t_2}\right)
\eea
for the underlying hyper-K$\ddot{\textrm{a}}$hler structure; for adding flavors, we allow the  $X^{4,5,6,11}$ part of the spacetime to be curved
with Taub-NUT singularities of collapsing $x^{11}$ circle.

Note how the central charge depends only on the homology class of the boundary 1-cycle.
Both $t_{1,2}$ live in $\IC^*$ whose two phase variables span two 1-cycles,
independent of each other. Decomposing
\bea
\partial \textrm{M2} =\sum_A \gamma_A
\eea
with $\gamma_A$ being the connected components of the boundary, the
Stoke's theorem gives
\begin{equation}
	Z_{\textrm{BPS}} =\sum \int_{\gamma_A} \lambda^{(A)}_{\textrm{SW}},
\end{equation}
where the differential $\lambda^{(A)}_{\textrm{SW}}$ is constrained to
be tangent to ${\gamma_A}$. If $\gamma_A$ carries a winding number along the
phase of $t_1$ but involves no winding along that of $t_2$, we must use
$\lambda^{(A)}_{\textrm{SW}} \sim \log(t_2) dt_1/t_1$, and vice versa. For the
more familiar 4d Seiberg-Witten theory, in contrast, one universal version
$\lambda\sim v\,dt/t$ worked for all BPS states. In the 5d setting, therefore,
individual $\lambda^{(A)}_{\textrm{SW}}$ to be used depends on the boundary
component; This subtle aspect will play an important role in the later
sections of this note.

In the following, we will mainly focus on the Coulomb phase of the 5d $SU(N)_k$ theory with Chern-Simons level $0 \leq k \leq N$. The Coulomb moduli space for $SU(N)$ theory is parametrized by the expectation values of the real adjoint scalar $\Phi$
\begin{equation}
	\Phi = \textrm{diag} \{ \phi_1 , \phi_2 , \cdots , \phi_N \},\quad \sum_{a=1}^N \phi_a = 0,
\end{equation}
where one choose a Weyl chamber of $SU(N)$ such that $\phi_1 > \phi_2 > \cdots > \phi_N$. The Cartan generators $\{H_a\}$ are the standard choices which read
\begin{equation}
	(H_a)_{bc} = \delta_{a,b} \delta_{a,c} - \delta_{a+1,b} \delta_{a+1,c},\quad a=1,\cdots,N-1.
\end{equation}
In the following we will take
\begin{equation}\label{eq-phi-varphi}
	\phi_a = \varphi_a - \varphi_{a-1},\quad a=1,\cdots,N,\quad \textrm{with}\quad \varphi_0=\varphi_N=0,
\end{equation}
where $\{\varphi_a\}$ are the analogue of the 5d central charges under the Cartan subgroup $U(1)^{N-1}$.

The spectral curve for $SU(N)$ theory can be read according to the rules in \eqref{Eq-mirror-curve}
\begin{equation}
	P_{SU(N)_k} = f_k(\lambda) \left(t_2^m t_1 + t_2^n t_1^{-1} \right) + \left(t_2^N + U_1 t_2^{N-1} + U_2 t_2^{N-2} + \cdots +U_{N-1} t_2 + 1  \right),
\end{equation}
which describes the 5d $SU(N)$ theory with Chern-Simons level $k$ compactified on a circle. The level $k$ is given by \cite{Closset:2018bjz}
\begin{equation}
	k = m + n - N,\quad (0\le m,n \le N)
\end{equation}
where $m$ and $n$ are the $y$-coordinates of the nodes $D_x$ and $D_y$ in the toric diagram. The factor $f_k(\lambda)$ is,
\begin{equation}\label{eq-f-lambda}
f_k(\lambda) = \left\{ \begin{array}{l}
\frac{1}{\lambda^{-\frac{1}{2}} + \lambda^{\frac{1}{2}}}\quad (k=\pm N) \\
\sqrt{\lambda}\quad (\textrm{others})
\end{array}\right.
\end{equation}
which can be read from the asymptotic behaviour $t_2\rightarrow 0,\infty$ of the brane web. The bare 5d instanton mass ${\mu}_0 \equiv 8\pi^2 / g_5^2$ is associated to $\lambda$ as
\begin{equation}
	\lambda = e^{-2\pi R_5 \mu_0},
\end{equation}
and $g_5$ is the 5d bare coupling. In particular, when $|\lambda|\ll 1$ and the effective 4d bare coupling is small, one can set $f_k(\lambda) = \sqrt{\lambda}$ for all $k$. The other parameters $\{U_a\}$ are related to the 5d Coulomb moduli such that in the asymptotic region of the K$\ddot{\textrm{a}}$hler moduli space one has
\begin{equation}
	U_a \approx e^{-2\pi R_5 a_a},
\end{equation}
where $\textrm{Re}(a_a) = \varphi_a$ are the complex versions of $\varphi_a$.

These spectral curves admit, collectively, a symmetry generated by
$\mathbb{Z}_{\textrm{gcd}(k,N)}$
\begin{equation}
	\mathbb{Z}_{\textrm{gcd}(k,N)} = \{1, g^{l}, g^{2l},\cdots, g^{(\textrm{gcd}(k,N)-1)l} \},
\end{equation}
with $l \equiv N/ \textrm{gcd}(k,N)$.
Consider the $\mathbb{Z}_N$ transformation $g$ which acts on the moduli
\begin{equation}
	U_a \xrightarrow{g} U_a e^{2\pi i \frac{a}{N}},\quad(a=1,\cdots,N-1)
\end{equation}
and combined with a redefinition of $t_1,t_2$ which reads
\begin{equation}
\quad t_2 \xrightarrow{g} t_2 e^{\frac{2\pi i}{N}},\quad t_1 \xrightarrow{g} t_1 e^{\frac{n}{N} 2\pi i},
\end{equation}
the polynomial $P(t_1,t_2)$ becomes
\begin{equation}
	P_g(t_1,t_2) = \sqrt{\lambda} \left(e^{\frac{k}{N}2\pi i} t_2^m t_1 + t_2^n t_1^{-1} \right) + \left(t_2^N + U_1 t_2^{N-1}  + \cdots +U_{N-1} t_2 + 1  \right).
\end{equation}
The first term $\sim t_2^m t_1$ comes with an addition phase factor $e^{\frac{k}{N} 2\pi i}$ under the transformation, where $k$ is the Chern-Simons level and we have used the relation $k = m + n - N$. As such, after $l$ repetitions of $g$,
the phase cancels out. In the special case $k=0$, $l=1$ and the group $\mathbb{Z}_N$ is
nothing but the center of $SU(N)$ generated by $g$. This $\mathbb{Z}_{\textrm{gcd}(k,N)}$
shuffles different holonomy saddles on the Coulomb moduli space among themselves, by
the left multiplication on $U$, whose phase parts are the $R_5$ holonomies
\begin{equation}
	\textrm{Arg}(U_a) \sim  \int_{\mathbb{S}^1} A^a \equiv h_a \ .
\end{equation}


\subsection{4d Holonomy Saddles}

As we have discussed, we use the word ``holonomy saddle" in the current context
to mean the 4d corners of this continuous Seiberg-Witten moduli space
where the resulting 4d theories yield interacting non-Abelian Seiberg-Witten
theories.
For example, for F0 theory one has two identical 4d Seiberg-Witten limits corresponding to $R_5 \langle A_5\rangle = 0$ and $R_5 \langle A_5\rangle = \sigma_3/2$, and they are related by a $\mathbb{Z}_2$ symmetry described above. In general, for $SU(N)$ theory one has $N$ holonomy saddles, and if Chern-Simons level $k$ is zero we have a $\mathbb{Z}_N$ symmetry relating such saddles. For non-zero $k$ the local 4d $\theta$-angle of the two adjacent saddles will differ by a shift of $2\pi k/N$ which will be shown in the following examples. Therefore the $\mathbb{Z}_N$ symmetry is broken to $\mathbb{Z}_{\textrm{gcd}(k,N)}$ in these cases.

\begin{figure}[pbth]
\centering
\includegraphics[scale=0.35]{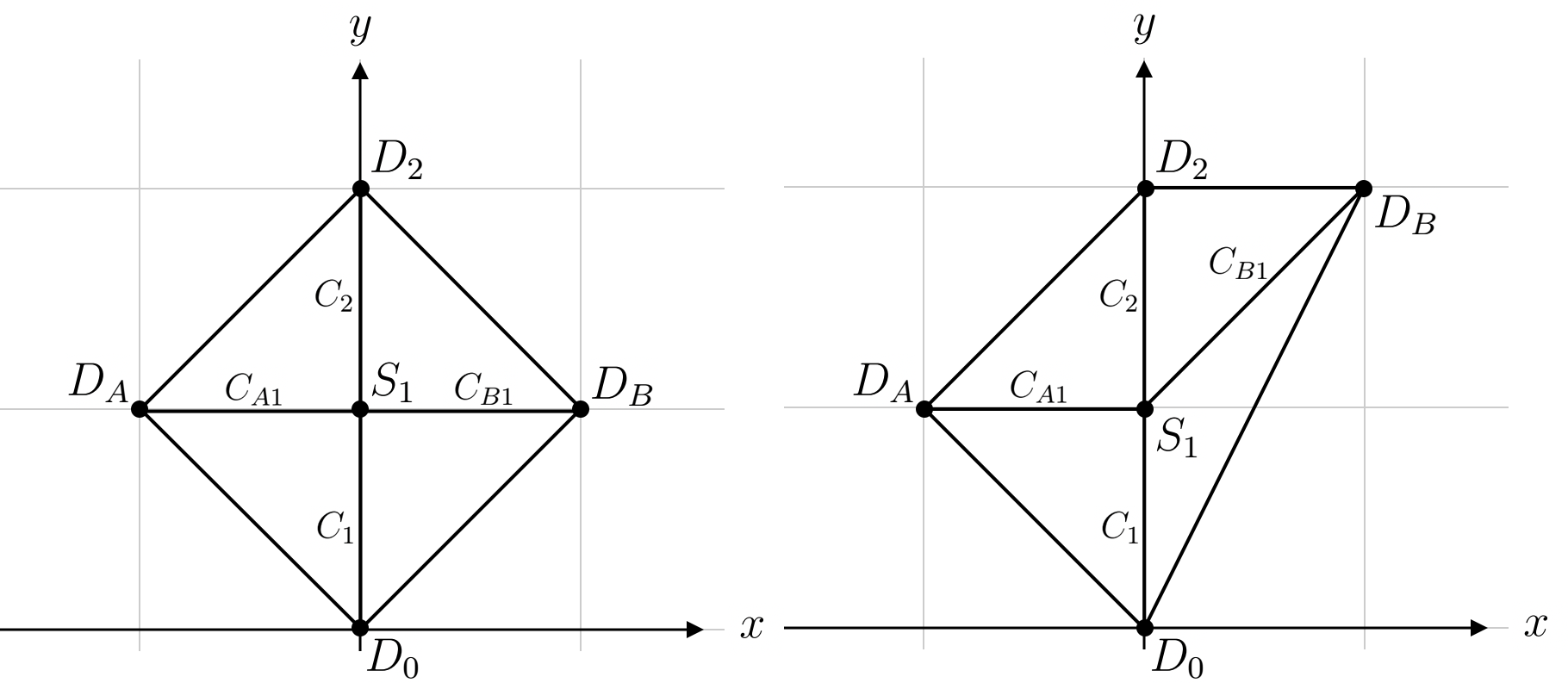}
\caption{\small The toric diagrams for $SU(2)_0$ (left) and $SU(2)_{\pi}$ (right) theories. Note how the diagrams
have the same general shape as $SU(N)_k$ of Fig. 1. This can be traced to the fact that the discrete theta
angle for $n=2$ is essentially the same thing as $k$ mod 2 since the relevant eta invariant takes discrete
values.}
\label{Fig-5dSU20-5dSU21-toric}
\end{figure}

Recall two kinds of $SU(2)$ theories differing by the discrete $\theta$-angle:
$SU(2)_{0}$ and $SU(2)_{\pi}$. They are separately engineered by compactfying
M-theory on F0 and F1 geometries whose toric diagrams can be read directly from \ref{Fig-toric-grid}.
Their spectral curves are, respectively
\begin{equation}\label{eq-5dSU20-mirror-curve}
	P_{F_0}(t_1,t_2) = \sqrt{\lambda} \left(\frac{t_2}{t_1} + t_1 t_2 \right) + \left(t_2^2 + U t_2 + 1 \right) = 0,
\end{equation}
for F0 theory and
\begin{equation}\label{eq-5dSU21-mirror-curve}
	P_{F_1}(t_1,t_2) = \sqrt{\lambda} \left(\frac{t_2}{t_1} + t_1 t_2^2 \right) + \left(t_2^2 + U t_2 + 1 \right) = 0,
\end{equation}
for F1 theory. The F0 curve admits the above $\mathbb{Z}_2$  ``symmetry", involving $U\rightarrow -U$.

There are two holonomy saddles labeled by $h=0,1$ on the $U$-plane located at $U = (-1)^{h} 2$. One may consider the local expansion
\begin{equation}
	U = (-1)^h (-2 + (2\pi R_5)^2 u),\quad t_2 = (-1)^h (-1+ (2\pi R_{5})z),
\end{equation}
where and the two polynomials $P_{F_0}(t_1,t_2)$ and $P_{F_1}(t_1,t_2)$ can be reduced to,
\bea
P_{F_0}(t_1,t_2) &= &\sqrt{\lambda} (-1)^h (t_1^{-1} + t_1 + \mathcal{O}(2\pi R_5)) + (2\pi R_5)^2 (z^2 - u +  \mathcal{O}(2\pi R_5))\cr\cr
P_{F_1}(t_1,t_2) &=& \sqrt{\lambda} ((-1)^h t_1^{-1} - t_1 + \mathcal{O}(2\pi R_5)) + (2\pi R_5)^2 (z^2 - u +  \mathcal{O}(2\pi R_5)).
\eea
The goal is to rewrite the 5d polynomials $P$ as
\bea
P(t_1,t_2,U;\lambda )= (2\pi R_5)^2 P_{\rm 4d}(t,z,u;\Lambda_{\rm 4d}) +O((2\pi R_5)^3)
\eea
in the small $R_5$ limit, for some choice of $t$ and a 4d QCD scale $\Lambda_{\rm 4d}$.

We introduce $t$ variable
\begin{equation}
	t_1 = \frac{(2\pi R_5)^2 t}{(-1)^h\sqrt{\lambda}},
\end{equation}
for F0, while for F1-theory
\begin{equation}
	t_1 = -\frac{(2\pi R_5)^2 t}{\sqrt{\lambda}}, 
\end{equation}
Then with the complex QCD scale $\Lambda_{k,h}$'s
\begin{equation}
	\Lambda^4_{k,h} = (-1)^{kh} \Lambda^4 \ ,\qquad \lambda = (2\pi R_5\Lambda)^4,
\end{equation}
where $k=0,1$ labels  $SU(2)_0$
and $SU(2)_{\pi}$ theories. We fix $\Lambda_{k,h}$ and send the compactification
radius $R_5$ to zero to obtain
\begin{equation}
P_{SU(2)_{4d,k,h}} =\left( \frac{\Lambda^4_{k,h}}{t} + t\right) + z^2 - u ,
\end{equation}
so that the 4d Seiberg-Witten curve becomes $P_{SU(2)_{4d,k,h}}=0$.

Recall the QCD scale $\Lambda_{k,h}$ is related to the local $\theta$-angle
of the 4d SU(2) theory via the $\beta$-function as
\begin{equation}
	(\Lambda_{k,h})^4 \sim \frac{1}{(R_5)^4} e^{-\frac{8 \pi^2}{g^2_{4d}} + i \theta_{k,h}},
\end{equation}
given how $1/R_5$ naturally plays the role of the UV cut-off in the 4d sense.
Therefore for $SU(2)_0$ theory the local 4d theories at the two holonomy saddles
have the same $\theta$-angle reflecting the $\mathbb{Z}_2$ symmetry. On the
contrary for $SU(2)_{\pi}$ theory the $\theta$-angle at the two saddles differ
by $\pi$ and the $\mathbb{Z}_2$ symmetry is lost.\footnote{The overall and
continuous phase of $\Lambda$ arises from a (parameter) superpartner of $\lambda$
that becomes available upon the compactification. We will ignore this angle as it plays
no particular role in our discussion.}

For pure $SU(3)_k$, the toric diagrams can also be found in Fig.~\ref{Fig-toric-grid}, and the
spectral curve is
\begin{equation}
	P_{SU(3)_k} = f_k(\lambda) \left(\frac{t_2^m}{t_1} + t_1 t_2^n \right) + \left(t_2^3 + U_1 t_2^2 + U_2 t_2 + 1 \right) =0,
\end{equation}
where the level $k$ is given by $k = m+n-3$. As discussed before, $SU(3)_0$ and $SU(3)_3$ theories have a $\mathbb{Z}_3$ symmetry which acts as $U_1\rightarrow U_1 e^{\frac{2\pi i}{3}},\quad U_2 \rightarrow U_2 e^{\frac{4\pi i}{3}}$.
There are three holonomy saddles located at
\begin{equation}
	U_1 = 3 e^{\frac{2\pi i}{3} h}, \quad U_2 = 3 e^{\frac{4\pi i}{3} h}.
\end{equation}
labeled by $h=0,1,2$

At each holonomy saddle, we can reduce the spectral curve to the 4d $SU(3)$ Seiberg-Witten curve via the following local expansion,
\begin{align}
	U_1 &= e^{\frac{2\pi i}{3} h}\left( 3 + (2\pi R_{5})^2 u - (2\pi R_5)^3 v/2 \right) \\
	U_2 &= e^{\frac{4\pi i}{3} h}\left( 3 + (2\pi R_5)^2 u + (2\pi R_5)^3 v/2 \right) \\
	t_2 &= e^{\frac{2\pi i}{3} h} \left(-1 + (2\pi R_5)z \right)
\end{align}
such that the polynomial $P_{SU(3)_k} $ can be expanded as
\begin{equation}
P_{SU(3)_{k,h}} = (2\pi R_{5})^3\left( \left(\frac{\Lambda_{k,h}^6}{t} + t \right) +  z^3 - u z - v \right) + \mathcal{O}((2\pi R_{5})^4),
\end{equation}
in small $R_5$ as before, with
\begin{equation}
	t_1 = \frac{t}{(-1)^n e^{\frac{2\pi i n h}{3}} \sqrt{\lambda}},
\end{equation}
and
\begin{equation}
 \Lambda_{k,h}^6 = (-1)^{k+1} e^{\frac{2\pi i k h}{3}} \Lambda^6\ , \qquad \lambda = (2\pi R_5 \Lambda)^6 \ ,	
\end{equation}
in line with the $SU(2)$ examples earlier.

This results in 4d spectral curve $P_{SU(3)_{4d,k,h}}=0$ with
\begin{equation}
P_{SU(3)_{4d,k,h}} =  \left(\frac{\Lambda_{k,h}^6}{t} + t \right) +  z^3 - u z - v \ .
\end{equation}
Again, the QCD scale $\Lambda_{k,h}$ is related to the local $\theta$-angle of the 4d $SU(3)$ theory via the $\beta$-function as
\begin{equation}
	(\Lambda_{k,h})^6 \sim \frac{1}{(R_5)^5}\,e^{-\frac{8 \pi^2}{g^2_{4d}} + i \theta_{k,h}}.
\end{equation}
For $k=0,3$, the local $\theta$-angles at the three holonomy saddles are the same, reflecting
the $\mathbb{Z}_3$ symmetry. For $k=1,2$, the local $\theta$-angles between the two
adjacent saddles differ by $\pm 2\pi /3$ and the $\mathbb{Z}_3$ symmetry is lost.

\section{Gluing 4d $SU(2)$  Seiberg-Witten Saddles}

In this section, we will derive the quiver diagrams for 5d $SU(2)$ theories with the help of the spectral curve.
We begin with the simplest rank one $SU(2)_0$ theory and discuss the procedure of the construction in much detail, and move on to  $SU(2)_{\pi}$ and highlight the difference. What we observe in this minimal example should apply to $SU(3)_k$ straightforwardly, again because the discrete theta angle $\theta$ of $SU(2)$ plays the role of $k\neq 0$ of $SU(N>2)$
for many purposes. A detailed discussion of $SU(3)_k$ will follow in Sec. 4.

\subsection{$SU(2)_0$}

The spectral curve of F0-theory is given by \eqref{eq-5dSU20-mirror-curve}
\begin{equation}
\sqrt{\lambda} \left(\frac{t_2}{t_1} + t_1 t_2  \right) + \left(t_2^2 + U t_2 + 1 \right) = 0,
\end{equation}
which is also geometrically	 depicted in Fig.~\ref{Fig-5dSU20-brane-web}.
Let's work on the $t_2$ plane, by solving $t_1$ as
\begin{equation}
	t_1 = \frac{-(1+U t_2 + t_2^2)\pm \sqrt{(1+U t_2 + t_2^2)^2 - 4 \lambda t_2^2}}{2t_2 \sqrt{\lambda}} \ .
\end{equation}
We have two sheets of $t_2$-plane depending on the sign factor, connected via the
two square-root branch cuts on the $t_2$-plane determined by the zero loci of the discriminant.
In the following, we choose the plus sign in the solution of $t_1$ which amounts to
work with the left half of the spectral curve depicted in Fig.~\ref{Fig-5dSU20-brane-web}.

\begin{figure}[pbth]
\centering
\includegraphics[scale=0.6]{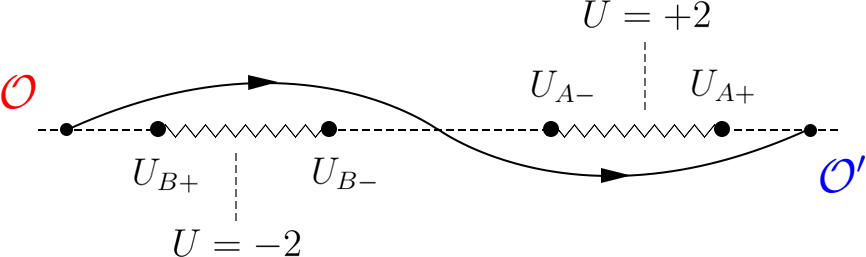}
\caption{\small The path in the $U$-plane connecting two observers $\mathcal{O}$ and $\mathcal{O}'$ probing the two holonomy saddles for $SU(2)_0$ theory.}
\label{Fig-5dSU20-U-plane-path}
\end{figure}

There are two holonomy saddles, located near $U=\pm 2$ on the $U$-plane as shown in Fig.~\ref{Fig-5dSU20-U-plane-path} and around each saddle there is a pair of singularities. We will stick to the convention in appendix A and call the half plane $\textrm{Re}(U)<0$ region $B$ and $\textrm{Re}(U)>0$ region $A$. The singularities at the two regions are also denoted as $U_{B\pm}$ and $U_{A\pm}$. We will label the two locations $\mathcal{O}$ and $\mathcal{O}'$ sitting at the real axis on the $U$-plane moduli space as shown in Fig.~\ref{Fig-5dSU20-U-plane-path}, symmetric under $U\rightarrow -U$.
Both of them would think the local theory is the 4d $SU(2)$ Seiberg-Witten theory.

The Seiberg-Witten differential $\lambda_{\textrm{SW}}$ is given by \eqref{eq-Seiberg-Witten-differential}
\begin{equation}\label{eq-Seiberg-Witten-differential-2}
	d \lambda_{\textrm{SW}} = \frac{1}{i(2\pi)^2 R_5 } \frac{d t_1}{t_1}\wedge \frac{d t_2}{t_2},
\end{equation}
The normalization factor is such that it will reduce to the 4d $\lambda_{\textrm{SW}}$ correctly at each holonomy saddle.
As pointed out in the previous section, in 5d Seiberg-Witten,
$\lambda_{\textrm{SW}}$ cannot be chosen universally but depends on the M2-brane configuration,
or more precisely on the topology of each boundary component thereof. {}From
\eqref{eq-Seiberg-Witten-differential-2} one finds $\lambda_{\textrm{SW}}$ can
be written as a linear combination in general
\begin{equation}
	i(2\pi)^2 R_5 \lambda_{\textrm{SW}} = c \log t_1 \frac{d t_2}{t_2} - (1-c) \log t_2 \frac{d t_1}{t_1},
\end{equation}For example, for the W-boson
one should use
$i(2\pi)^2 R_5 \lambda_{\textrm{SW}} = -(\log t_2) d t_1 /t_1 $ and for the
instanton particles $i(2\pi)^2 R_5 \lambda_{\textrm{SW}} = (\log t_1) d t_2 /t_2 $.

The BPS states are represented by the integral of a carefully chosen $\lambda_{\textrm{SW}}$
along the 1-cycles on the spectral curve and the 1-cycles will also deform as the observer moves on the $U$-plane moduli space.
Take the $\lambda_{\textrm{SW}}$ relevant for the W-boson. Near the two saddles we can rewrite
$t_1,t_2,U,\lambda$ in terms of the local 4d parameters $t,z,u,\Lambda$ following the discussion before and one finds
\begin{equation}\label{Eq-5dSU20-local-SW-differential}
\lambda_{\textrm{SW}} \approx \frac{1}{2\pi i} \frac{z d t}{t}\quad (\textrm{At} \ U=-2),\quad \lambda_{\textrm{SW}} \approx \frac{1}{2\pi i} \frac{z d t}{t}\pm \frac{d t}{4 \pi R_5 t} \quad (\textrm{At} \ U=2).
\end{equation}
Near $U=-2$ the Seiberg-Witten differential $\lambda_{\textrm{SW}}$ reduce to exactly the 4d $\lambda_{\textrm{SW}}$ for a pure $SU(2)$ Seiberg-Witten theory. However, near the other saddle $U=2$ there is an additional piece given by a total derivative $dt/(4\pi R_5 t)$ and for any 1-cycle carrying a winding number of $t$, it will contribute half the KK-charge\footnote{This additional piece cannot be avoided. If one tries to eliminate it by shifting $\lambda_{\textrm{SW}}$ with the same total derivative, it will then show up in the 4d $\lambda_{\textrm{SW}}$ at $U=-2$ instead.
This term is responsible for the KK-charge associated with one of the quiver nodes in the 5d quiver diagram.}. This term comes from the phase of $\log t_2$ and the sign factor depends on how you move to $U=2$ from $U=-2$. We will fix the sign later in an explicit example.

It is also important to note that there are two other singularities of $\lambda_{\textrm{SW}}$
at $t_2=0,\infty$ on the $t_2$ plane. They correspond to the two asymptotic 1-cycles drawn in Fig.~\ref{Fig-5dSU20-brane-web} and
contribute to the instanton charge. To evaluate them one needs to choose $\lambda_{\textrm{SW}}$
correctly and it turns out that each 1-cycle will contribute half of the instanton central charge $\mu_0/2$.
Together with how KK charge contribution arises, as described above, this gives us the
basic mechanism of how, starting with 4d BPS objects, we end up constructing 5d BPS states
simply by moving from one saddle to the other. By doing things backward, i.e., by considering
4d BPS states in the other saddle and bringing them back to the first saddle, we end up
collecting all 4 BPS states in the first saddle, which will eventually span the 5d BPS
quiver of a pure $SU(2)$ theory.


\begin{figure}[pbth]
\centering
\includegraphics[scale=0.5]{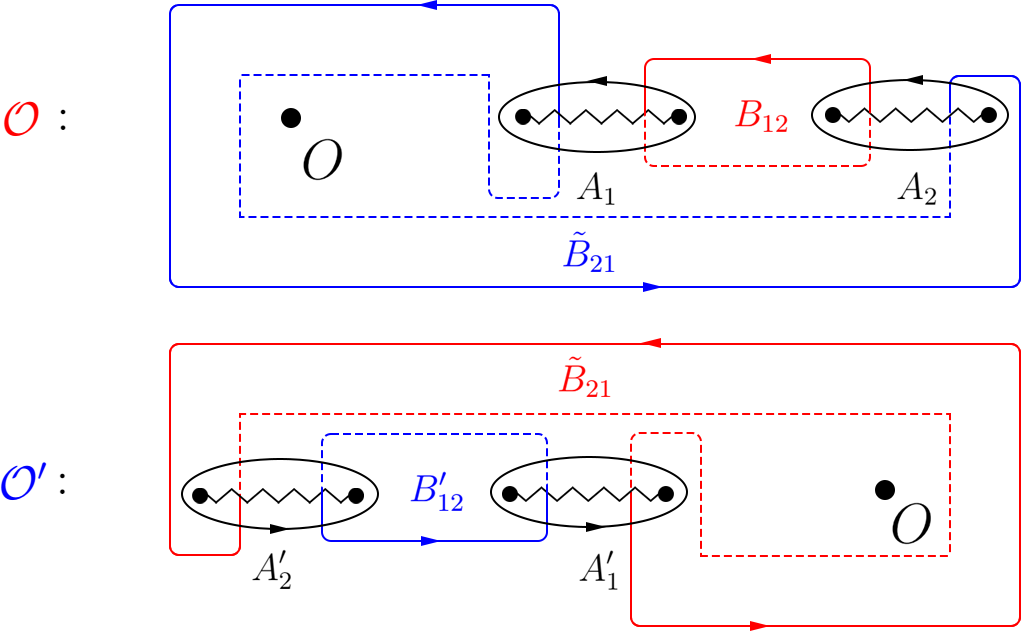}
\caption{\small The local branch cuts and choices of 1-cycles on $t_2$-plane for the observers $\mathcal{O}$ and $\mathcal{O}'$.}
\label{Fig-5dSU20-curve-cycle}
\end{figure}

\begin{figure}[pbth]
\centering
\includegraphics[scale=0.6]{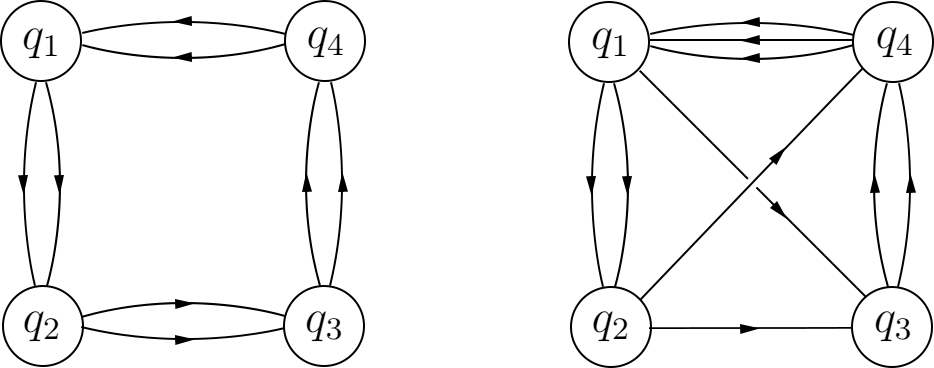}
\caption{\small The quiver diagrams for 5d $SU(2)_0$ theory (left) and $SU(2)_{\pi}$ theory (right).}
\label{Fig-5dSU2-quiver-all}
\end{figure}

Note that
with positive $\lambda$, the branch cuts on $t_2$-plane will lie on the real axis and are drawn in Fig.~\ref{Fig-5dSU20-curve-cycle} for observers $\mathcal{O}$ and $\mathcal{O}'$.
Let's first consider the saddle $U=-2$ probed by $\mathcal{O}$.  We can choose a standard basis of 1-cycles on the $t_2$-plane as $A_1,A_2,B_{12}$ shown in the figure. Starting at $\mathcal{O}$, if we approach the two singularities $U_{B\pm}$ in Fig.~\ref{Fig-5dSU20-U-plane-path} form the second quadrant, one will find the $B_{12}$ collapse at $U_{B+}$ and $ -B_{12} + A_1 - A_2$ collapse at $U_{B-}$. Therefore these two contours correspond to the  $(1,0)$ monopole and $(-1,2)$ dyon of the local 4d Seiberg-Witten theory and are represented by the left two nodes of the first quiver diagram in Fig.~\ref{Fig-5dSU2-quiver-all}.

We also need to find the states represented by the other two nodes. Apparently, they should come from the other holonomy saddle probing by $\mathcal{O}'$ where the branch cuts and the choices of 1-cycles $A'_1,A'_2,B'_{12}$ are shown in the second diagram in Fig.~\ref{Fig-5dSU20-U-plane-path}. However, the quiver diagram is drawn at a specific point on the moduli space. That means one of the observers, let's say observer $\mathcal{O}'$, must move to the other observer $\mathcal{O}$ and combine their local 4d quivers to form the 5d quiver. In order to do so, we will connect the two holonomy saddles probing by $\mathcal{O}$ and $\mathcal{O}'$ on the $U$-plane in a symmetric way via a path through the origin depicted in Fig.~\ref{Fig-5dSU20-U-plane-path}.

The contour for one of the remaining two nodes is the blue contour $\tilde{B}_{21}$ shown in Fig.~\ref{Fig-5dSU20-curve-cycle} which encloses the origin twice\footnote{More precisely, the contours enclose the origin contains one contour (solid line) circling the origin on the first $t_2$-sheet, and another contour (dashed line) circling the origin on the second $t_2$-sheet reversely. Their contributions are actually the same.}. The residue of $\lambda_{\textrm{SW}}$ at the origin contributes to half the instanton charge, therefore this state must carry a single instanton charge. If we denote the solid contour circling the origin $O$ only (counter-clockwisely) as $\mathcal{C}_O$, then $\tilde{B}_{21}$ is decomposed as $\tilde{B}_{21} = -B_{12} + A_1 + A_2 + 2\mathcal{C}_O$. 

\begin{figure}[pbth]
\centering
\includegraphics[scale=0.35]{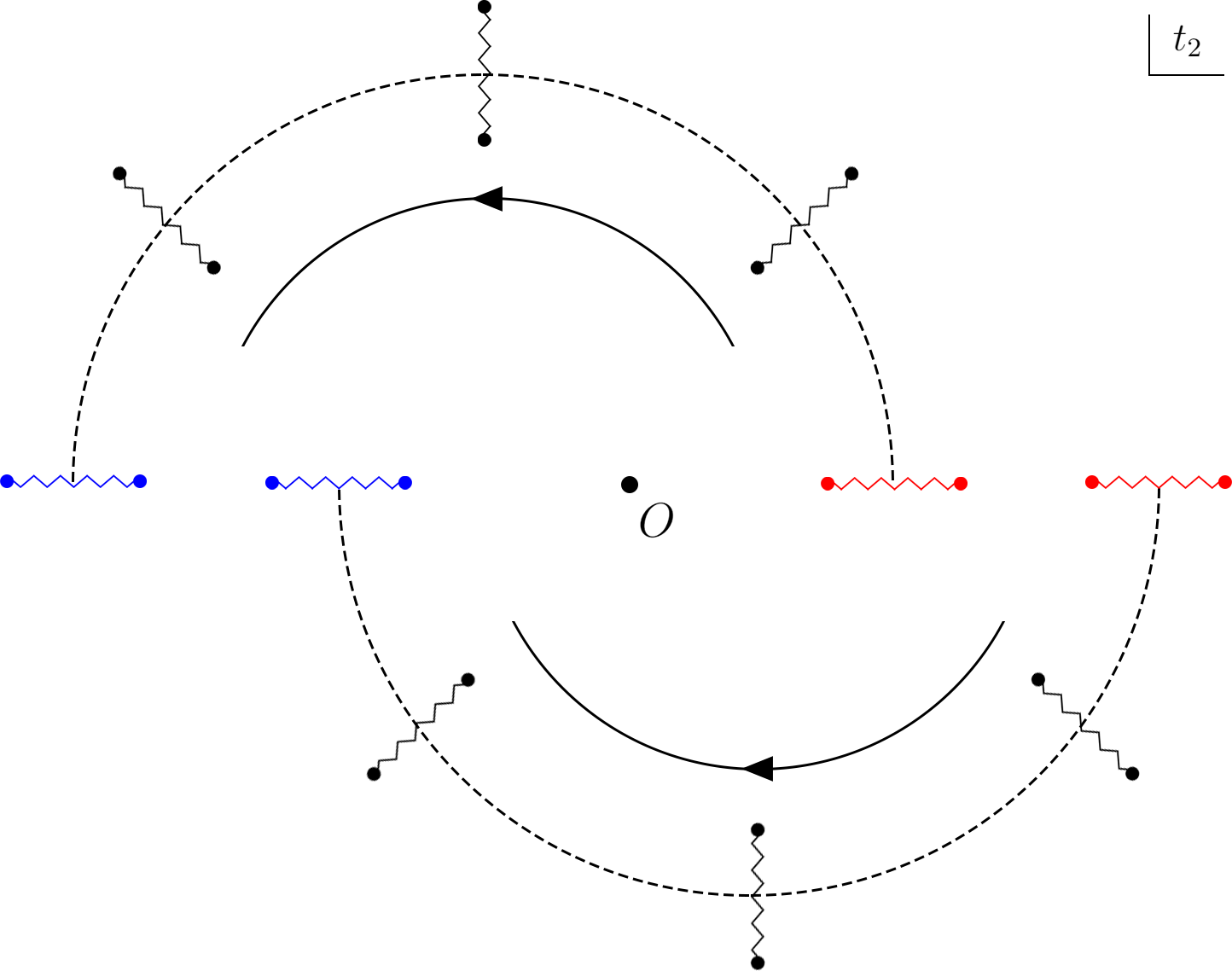}
\caption{\small The deformations of branch cuts on the spectral curve ($t_2$-plane) from $\mathcal{O}$ to $\mathcal{O}'$ on the $U$-plane for $SU(2)_0$ theory.}
\label{Fig-5dSU20-curve-cycle-rotation}
\end{figure}

Let us explain why this complicated contour $\tilde{B}_{21}$ is special.  Consider moving from $\mathcal{O}$ to $\mathcal{O}'$ along the path described in Fig.~\ref{Fig-5dSU20-U-plane-path}, the spectral curve will also deform since it depends on the moduli $U$. To be more explicit, we trace the deformations of the two branch cuts on the spectral curve and depict them in Fig.~\ref{Fig-5dSU20-curve-cycle-rotation}.
Initially, for $\mathcal{O}$ on the $U$-plane, the two branch cuts are colored in red and they will rotate following the dashed line to reach the blue branch cuts at the other side, which corresponds to $\mathcal{O}'$ on the $U$-plane. The $A$-cycles transform as
\begin{equation}\label{5dSU20-curve-cycle-transformation-A}
A_1 \rightarrow A'_2, \quad A_2 \rightarrow A'_1,
\end{equation}
where $A'_1,A'_2$ are shown in the second diagram in Fig.~\ref{Fig-5dSU20-curve-cycle}.

Similarly, if we trace the blue contour $\tilde{B}_{21}$ in the first diagram
in Fig.~\ref{Fig-5dSU20-curve-cycle}, we will find it becomes the simple blue contour $B'_{12}$ in the second diagram, namely
\begin{equation}\label{5dSU20-curve-cycle-transformation-B}
	\tilde{B}_{21} \rightarrow B'_{12}.
\end{equation}
On the other hand, the red contour $B_{12}$ will transform into the red contour $\tilde{B}'_{21}$ which becomes complicated instead. The local Seiberg-Witten differential at $U=2$ is shifted by a total differential $\pm dt/(4\pi R_5 t)$ as discussed in \eqref{Eq-5dSU20-local-SW-differential}. However, the $B'_{12}$ cycle does not carry any winding numbers of $t$ so this contour corresponds to the $(1,0)$ monopole of the second local 4d Seiberg-Witten theory probed by $\mathcal{O}'$.

Now we can simply write down the contour for the fourth node based on the previous discussion. This must be a complicated contour that becomes simple when we move to $\mathcal{O}'$ and we expect it to be the $(-1,2)$ dyon of the second local 4d Seiberg-Witten theory at $\mathcal{O}'$. Therefore the fourth contour should be $-\tilde{B}_{21} - A_1 + A_2$ such that if we push the contour to $\mathcal{O}'$ by applying the transformation \eqref{5dSU20-curve-cycle-transformation-A} and \eqref{5dSU20-curve-cycle-transformation-B} we get $-B'_{12} + A'_1 - A'_2$, which is associated to the $(-1,2)$ dyon of the second local 4d Seiberg-Witten theory at $\mathcal{O}'$. Moreover, the $A'$-cycles carry the winding number of $t$, therefore the total differential $\pm dt/(4\pi R_5 t)$ will contribute.

The sign is determined in the following way. The 5D Seiberg-Witten differential is given by $\lambda_{\textrm{SW}} = -(4\pi^2 R_5 i)^{-1} dt_1/t_1 \log t_2$ before reduction on the holonomy saddles. The $t_2$-plane is depicted in Fig.~\ref{Fig-5dSU20-curve-cycle-rotation} and $A_1,A_2$ cycles rotate to $A'_2,A'_1$ in different way as shown in the figure. During the rotation the $\log t_2$ will pick up a phase $-\pi i$ for $A'_1$ and $+\pi i$ for $A'_2$ and the total differential is $+dt/(4\pi R_5 t)$ for $A'_1$ and $-dt/(4\pi R_5 t)$ for $A'_2$. Therefore the contour $-B'_{12} + A'_1 - A'_2$ actually gives a $(-1,2)$ dyon with an addition KK-charge $i/R_5$. In order to get exactly the $(-1,2)$ dyon state we should start with the contour $\tilde{B}_{21} - A_1 + A_2 - [\textrm{KK}]$ for the fourth node, where $[\textrm{KK}]$ represents a contribution of KK-charge $i/R_5$.

As a summary, with the base point located at $\mathcal{O}$ we have four 1-cycles represented as $B_{12}$, $-B_{12}+A_1-A_2$, $\tilde{B}_{12}$ and $-\tilde{B}_{12}-A_1+A_2-[\textrm{KK}]$. Here $\tilde{B}_{12}$ can also be decomposed as
\begin{equation}
\tilde{B}_{21} = -B_{12} + A_1 + A_2 + 2\mathcal{C}_O.
\end{equation}
Actually, the combination of $A_1$ and $A_2$ is trivial due to the following reason. The contour enclosing $A_1$ and $A_2$ is equivalent to the reversed contour enclosing the origin $t_2=0$ and the infinity point $t_2 \rightarrow \infty$. One can check they cancel with each other and therefore the integral of $\lambda_{\textrm{SW}}$ along $A_1+A_2$ is zero.
One possible charge assignments for these four BPS states at base point $\cO$ are
\begin{itemize}
\item $B_{12}$ : A monopole of charge $(1,0,0,0)$
\item $-B_{12}+A_1-A_2$: A dyon of charge $(-1,2,0,0)$
\item $-B_{12}+2\mathcal{C}_O$: An anti-monopole of charge $(-1,0,0,1)$, carrying a unit of instanton charge
\item $B_{12} - A_1 + A_2 - 2\mathcal{C}_O-[\textrm{KK}]$: A dyon of charge $(1,-2,-1,-1)$, carrying $-1$-unit of both KK and  instanton charge
\end{itemize}
Moreover, the intersection numbers between the 1-cycles are chosen to be
\begin{equation}\label{eq-5dSU20-curve-cycle-intersection-number}
	A_1 \# B_{12} = B_{12} \# A_2 = 1,
\end{equation}
such that it is consistent with the Dirac pairing of the charges. Therefore we reproduce the 5d quiver diagram given by the left diagram in Fig.~\ref{Fig-5dSU2-quiver-all} \cite{Closset:2019juk}.

\vskip 5mm


\subsection{$SU(2)_{\pi}$}

With the above $SU(2)_0$ example, the strategy is clear: we construct two local 4d quivers at two saddles symmetrically as some contours on the $t_2$-plane, and then pull the contours at the second saddle back to the first saddle along a path on the moduli space. We will apply this procedure to $SU(2)_{\pi}$.

\begin{figure}[pbth]
\centering
\includegraphics[scale=0.6]{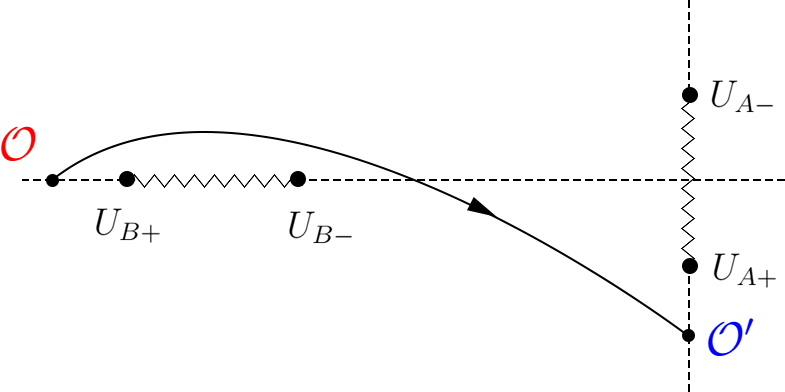}
\caption{\small The path in the $U$-plane connecting two observers $\mathcal{O}$ and $\mathcal{O}'$ probing the two holonomy saddles for $SU(2)_{\pi}$ theory.}
\label{Fig-5dSU21-U-plane}
\end{figure}

For F1 theory the spectral curve is given by \eqref{eq-5dSU21-mirror-curve} as
\begin{equation}
	P_{F_1}(t_1,t_2) = \sqrt{\lambda} \left(\frac{t_2}{t_1} + t_1 t_2^2 \right) + \left(t_2^2 + U t_2 + 1 \right) = 0,
\end{equation}
which does not possess the $\mathbb{Z}_2$ symmetry. Without loss of generality, we will adjust the phase of $\lambda$ such that the two singularities on the $U$-plane at the first holonomy saddle $U=-2$ are lying on the real axis, just like that in Fig.~\ref{Fig-5dSU20-U-plane-path} for $SU(2)_0$ theory. Then the other two singularities at the second holonomy saddle $U=2$ in Fig.~\ref{Fig-5dSU20-U-plane-path} are rotated by $90$ degree which is sketched in Fig.~\ref{Fig-5dSU21-U-plane}.

\begin{figure}[pbth]
\centering
\includegraphics[scale=0.35]{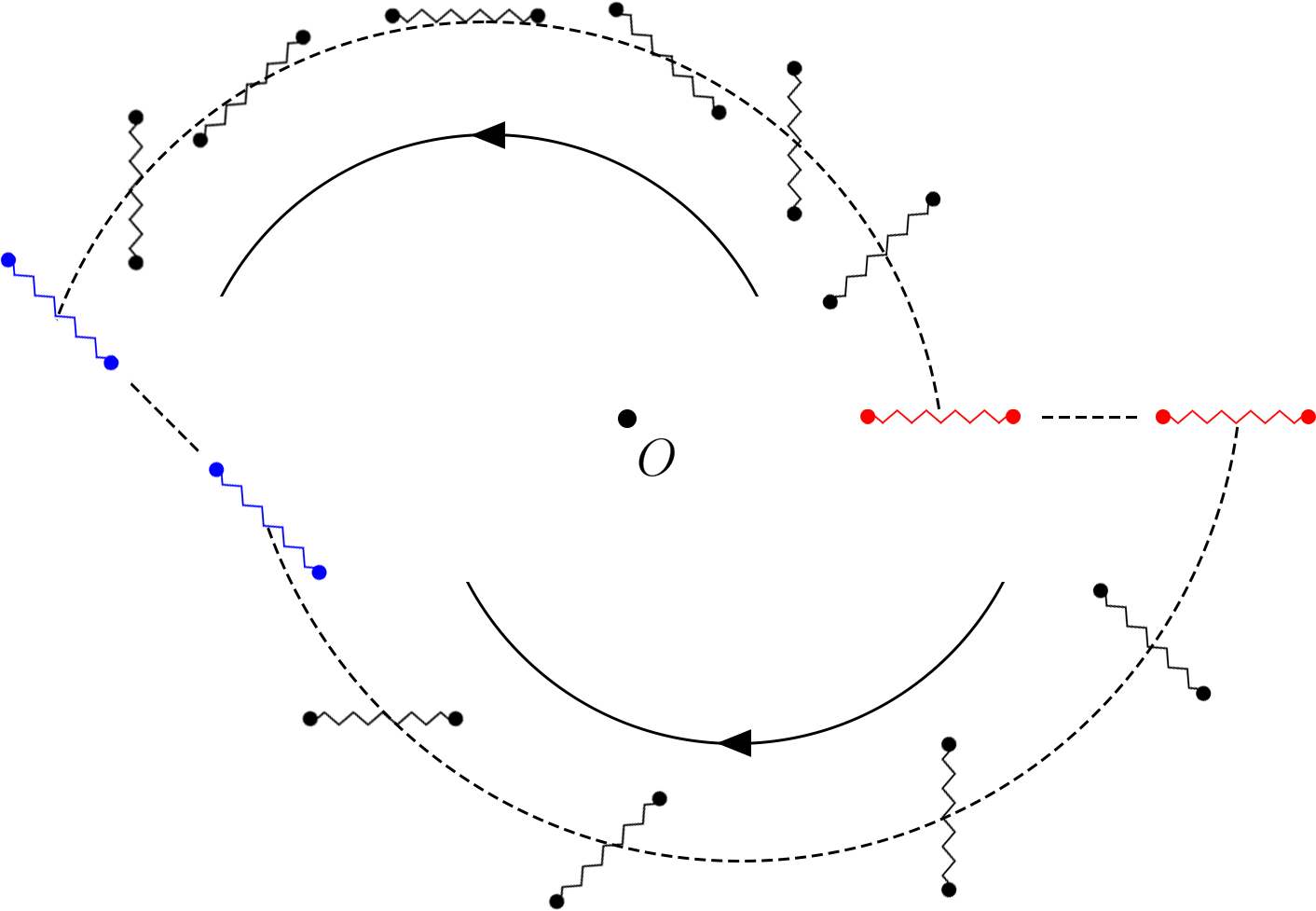}
\caption{\small The transformations of branch cuts of the $t_2$-plane following the path from $\mathcal{O}_B$ to $\mathcal{O}_A$ on the $U$-plane for $SU(2)_{\pi}$ theory.}
\label{Fig-5dSU21-curve-cycle-rotation-1}
\end{figure}

\begin{figure}
\centering
\includegraphics[scale=0.5]{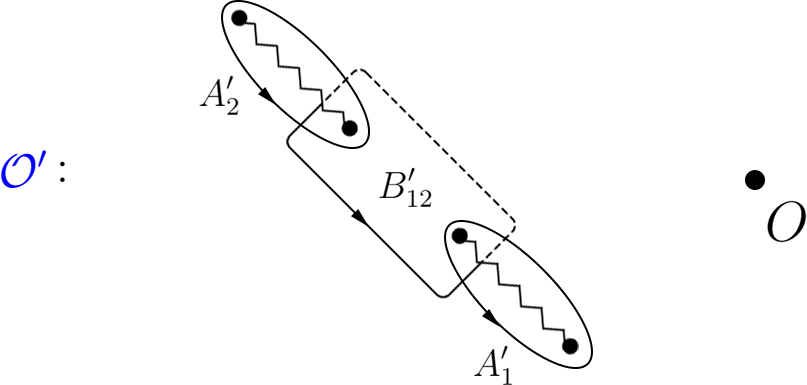}
\caption{\small The choices of 1-cycles for $SU(2)_{\pi}$ theory at the second holonomy saddle.}
\label{Fig-5dSU21-curve-cycle-OA}
\end{figure}

Following the previous discussion, let's consider the two observers $\mathcal{O}$ and $\mathcal{O}'$ as shown in Fig.~\ref{Fig-5dSU21-U-plane} and their local geometries are the same. Then we can follow the path described in Fig.~\ref{Fig-5dSU20-U-plane-path} and trace the transformation of the spectral curves. There is a complication for F1-theory: the branch cuts on the spectral curve for $\mathcal{O}'$ are no longer the blue cuts given in Fig.~\ref{Fig-5dSU20-curve-cycle-rotation}. This can be seen as follows. The local 4d curve is written as
\begin{equation}\label{eq-5dSU21-theta-angle-redefine}
	P_{SU(2)_{4d}} = \left(\frac{\Lambda^4}{t}+t\right) + z^2 - u=0,
\end{equation}
where $\Lambda^4$ is proportional to $e^{i \theta}$ and $\theta$ is the 4d $\theta$-angle. Shifting the $\theta$-angle by $\Delta$ is amount to multiplying $\Lambda^4$ by $e^{i \Delta}$. It can be compensated by the following transformation
\begin{equation}
t \rightarrow t e^{\frac{i \Delta}{2}},\quad z \rightarrow z e^{\frac{i \Delta}{4}},\quad u \rightarrow u e^{ \frac{i \Delta}{2}},
\end{equation}
which will bring the curve back to its original form. For the second holonomy saddle, the local $\theta$-angle is shifted by $\Delta = \pi$ as discussed before so that the local $u$-plane is rotated by 90-degree which gives vertical branch cuts in Fig.~\ref{Fig-5dSU21-U-plane} connecting the $U_{A+}$ and $U_{A-}$. Further, the two branch cuts on the spectral curve (parametrized by $z$ locally) will rotate a $45$-degree corresponding to the two slanted blue cuts in Fig.~\ref{Fig-5dSU21-curve-cycle-rotation-1}. The dash lines show how the branch cuts transform when we follow the path depicted in Fig.~\ref{Fig-5dSU21-U-plane}.
Since the local theories at $\mathcal{O}$ and $\mathcal{O}'$ are totally the same, one may similarly choose the 1-cycles at $\mathcal{O}'$ as shown in Fig.~\ref{Fig-5dSU21-curve-cycle-OA}.

\begin{figure}[pbth]
\centering
\includegraphics[scale=0.5]{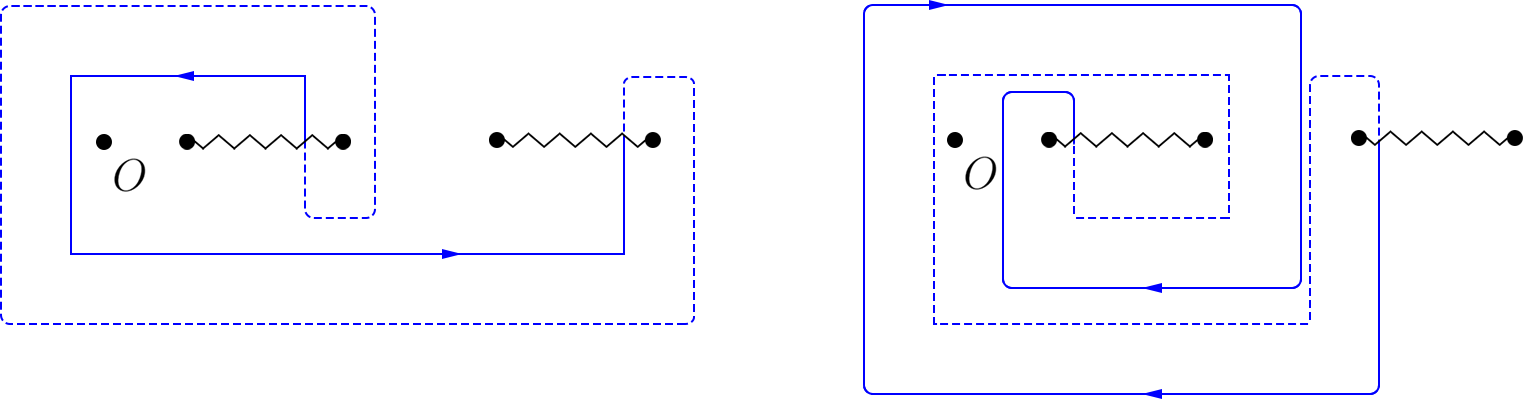}
\caption{\small The pullback of the contours $B'_{12}$ (left) and $-B'_{12} + A'_1 - A'_2$ (right) to $\mathcal{O}$.}
\label{Fig-5dSU21-curve-cycle-OA-to-OB}
\end{figure}

With the help of Fig.~\ref{Fig-5dSU21-curve-cycle-rotation-1} we can determine the 5d quiver diagram straightforwardly as we did in the $SU(2)_0$ case. We have two 4d theories located at $\mathcal{O}$ and $\mathcal{O}'$. For both of them the local quivers consist two nodes representing the $(1,0)$-monopole and $(-1,2)$-dyon, which are further associated with the contours $B_{12}$ and $-B_{12} + A_1 - A_2$ at $\mathcal{O}$ or $B'_{12}$ and $-B'_{12} + A'_1 - A'_2 - [\textrm{KK}]$ at $\mathcal{O}'$. Where the shift of KK-charge is due to the same reason discussed in the $SU(2)_0$ theory.  The next step is to combine these two 4d quivers to make the 5d quivers. In order to do so we will follow the transformation in Fig.~\ref{Fig-5dSU21-curve-cycle-rotation-1} to pull the contours $B'_{12}$ and $-B'_{12} + A'_1 - A'_2 - [\textrm{KK}]$ at $\mathcal{O}'$ back to $\mathcal{O}$. The results are shown in Fig.~\ref{Fig-5dSU21-curve-cycle-OA-to-OB}.
Both of them enclose the origin of $t_2$-plane twice which give rise to the instanton charges. They can be decomposed using the local basis $A_1,A_2,B_{12}$ at $\mathcal{O}$ as
\begin{equation}
	B'_{12}\rightarrow -B_{12} + A_1 + 2 \mathcal{C}_O,\quad -B'_{12} + A'_1 - A'_2 - [\textrm{KK}] \rightarrow B_{12} - 3A_1 - 2\mathcal{C}_O - [\textrm{KK}],
\end{equation}
where $\mathcal{C}_O$ is the contour circling the origin counter-clockwisely and we have used the fact that $A_1 + A_2$ is trivial.

At the base point $\mathcal{O}$, one possible charge assignments for the resulting four BPS states are
\begin{itemize}
\item $B_{12}$ : A monopole of charge $q_1=(1,0,0,0)$
\item $-B_{12}+2A_1$: A dyon of charge $q_2=(-1,2,0,0)$
\item $-B_{12}+A_1+2\mathcal{C}_O$: An dyon of charge $q_3=(-1,1,0,1)$, carrying one unit of instanton charge
\item $B_{12} - 3A_1 - 2\mathcal{C}_O - [\textrm{KK}]$: A dyon of charge $q_4=(1,-3,-1,-1)$, carrying $-1$-unit of instanton charge and KK charge
\end{itemize}
which are consistent with the intersection numbers \eqref{eq-5dSU20-curve-cycle-intersection-number}. These give the second quiver in Fig.~\ref{Fig-5dSU2-quiver-all}.

We may follow the same method described in the previous sections to obtain the 5d quivers for F2 theory. As before two nodes of the 5d quiver is given by the contours $B_{12}$ and $-B_{12} + A_1 - A_2$ at $\mathcal{O}$ and the other two nodes are $B'_{12}$ and $-B'_{12} + A'_1 - A'_2 - [\textrm{KK}]$ from the other saddle at $\mathcal{O}'$. We then need to pull them back to $\mathcal{O}$ following the path described in Fig.~\ref{Fig-5dSU20-U-plane-path}. The results are
\begin{equation}
	B'_{12}\rightarrow -B_{12} + 2A_1 + 2A_2 + 2\mathcal{C}_O,\quad -B'_{12} + A'_1 - A'_2 - [\textrm{KK}] \rightarrow B_{12} - 3 A_1 - A_2 - 2\mathcal{C}_O - [\textrm{KK}].
\end{equation}
Moreover since $A_1 + A_2$ is trivial, we get the same two contours as in the F0 theory. Therefore the 5d quiver diagram for F2 theory is the same to that for F0 theory as expected.

\section{5d $SU(3)_k$ BPS Quivers}

Let us move on to $SU(3)$. We will mainly repeat the previous exercise for  $SU(3)_0$ and $SU(3)_1$
 as an illustration and in the last subsection present results for $SU(3)_{2,3}$. The resulting BPS quivers
 are presented below in Fig.~\ref{Fig-5dSU3-quiver} \cite{Closset:2019juk}, each of which reproduces the standard D0  theory probing  the
 respective local Calabi-Yau.

\vskip 5mm
\begin{figure}[pbth]
\centering
\includegraphics[scale=0.4]{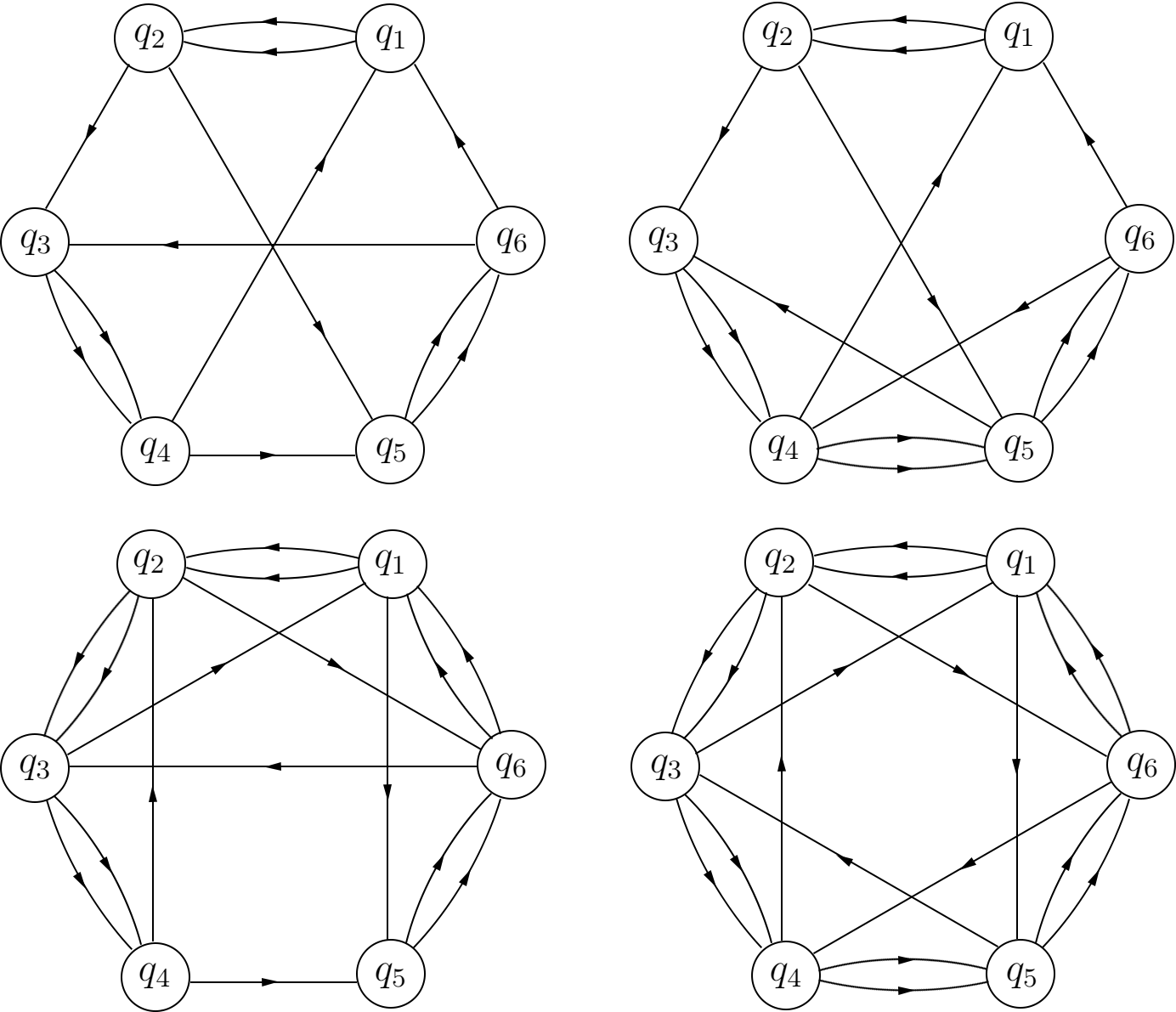}
\caption{\small The quiver diagram for 5d $SU(3)_k$ theories for $k=0,1,2,3$ (top-left, top-right, bottom-left and bottom-right). }
\label{Fig-5dSU3-quiver}
\end{figure}

\subsection{$SU(3)_0$}

We begin with $SU(3)_0$ theory whose spectral curve is given by
\begin{equation}
	P_{SU(3)_0} = \sqrt{\lambda} \left(\frac{t_2^2}{t_1} + t_1 t_2 \right) + \left(t_2^3 + U_1 t_2^2 + U_2 t_2 + 1 \right) = 0,
\end{equation}
with the $\mathbb{Z}_3$ symmetry generated by
\begin{equation}
U_1\rightarrow U_1 e^{\frac{2\pi i}{3}},\quad U_2 \rightarrow U_2 e^{\frac{4\pi i}{3}}.
\end{equation}
The three local $SU(3)$ holonomy saddles are identical and they permute under the $\mathbb{Z}_3$ transformation. At each saddle, the local theory is described by the 4d $SU(3)$ Seiberg-Witten theory and a reduction of the spectral curve is presented in section 2. Unlike $SU(2)$ theory, the moduli space for compactified $SU(3)$ theory is complex 2-dimensional parametrized by $U_1$ and $U_2$ and the 'singularities' will be (complex) co-dimensional one singular strings, which makes the analysis of the moduli space rather complicated. Nevertheless, we will work with a specific (complex) hyperplane in the following which is sufficient to derive the 5d quiver.

Let's consider a specific hyperplane $U_1 = U_2 = U$ and call it $U$-plane in the following. This suffices for explaining how we move from one saddle to the next. The spectral curve is simplified as
\begin{equation}
	P_{SU(3)_0} = \sqrt{\lambda} \left(\frac{t_2^2}{t_1} + t_1 t_2 \right) + \left(t_2^3 + U t_2^2 + U t_2 + 1 \right) = 0.
\end{equation}
There are three branch cuts on the $t_2$-plane solved by the discriminant equation
\begin{equation}
	\left(t_2^3 + U t_2^2 + U t_2 + 1 \right)^2 - 4 \lambda t_2^3 = 0,
\end{equation}
and the singularities on the $U$-plane are further determined by solving the discriminant of the above one with respect to $t_2$. One of the 4d holonomy saddles is located at $U=3$ and the local theory is like a 4d $SU(3)$ Seiberg-Witten theory.

\begin{figure}[pbth]
\centering
\includegraphics[scale=0.6]{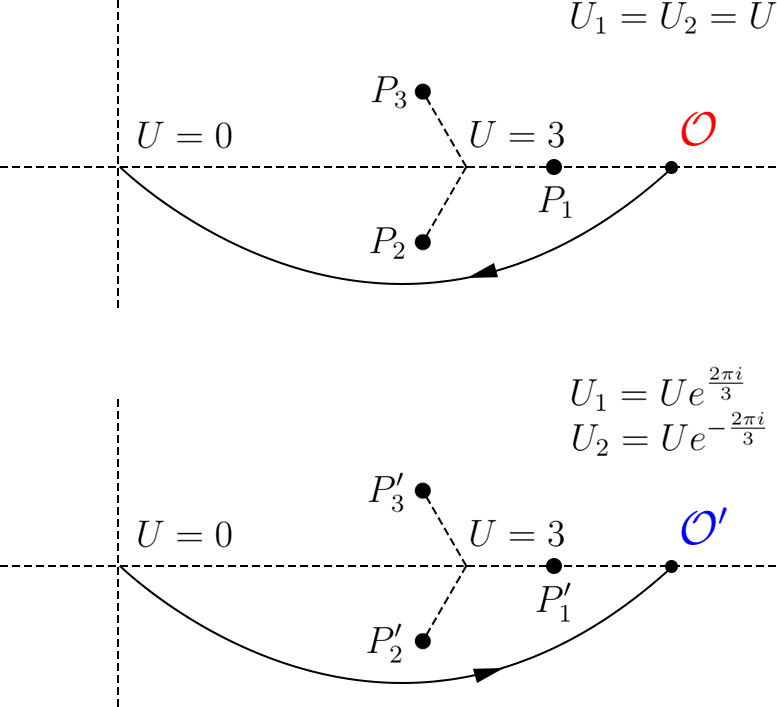}
\caption{\small The $U$-plane for $SU(3)_0$ theory and the path connecting the two observers $\mathcal{O}_B$ and $\mathcal{O}_A$ probing the two holonomy saddles. }
\label{Fig-5dSU30-U-plane}
\end{figure}

\begin{figure}[pbth]
\centering
\includegraphics[scale=0.5]{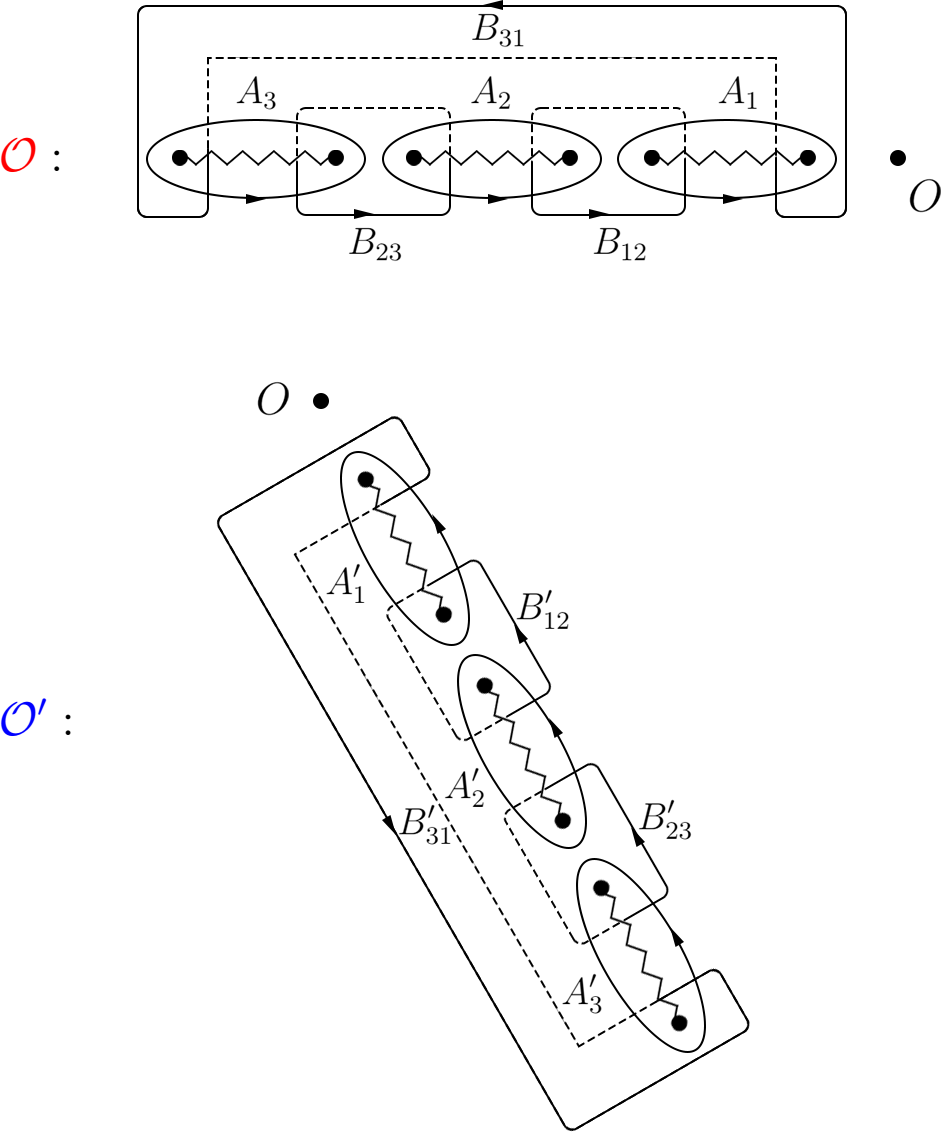}
\caption{\small The local branch cuts and choices of 1-cycles on $t_2$-plane for the observers $\mathcal{O}$.}
\label{Fig-5dSU30-curve-cycle}
\end{figure}

There are three singularities\footnote{On a generic (complex) hyperplane in the moduli space there are six singularities. However, with the special choice $U_1=U_2$ there will be three double-degenerated singularities.}
$P_1,P_2,P_3$ surrounding $U=3$ at the $U$-plane which are depicted in the first diagram in Fig.~\ref{Fig-5dSU30-U-plane}. Here we have adjusted the phase of $\lambda$ such that one of them will lie on the real axis.
We consider a base point located at $\mathcal{O}$ and the branch cuts on the spectral curve ($t_2$-plane) are simply shown in the first diagram in Fig.~\ref{Fig-5dSU30-curve-cycle},
where standard choices of the $A/B$-cycles are also depicted in the figure. The Seiberg-Witten differential $\lambda_{\textrm{SW}}$ is again chosen as $\lambda_{\textrm{SW}} = -(4\pi^2 R_5 i)^{-1} d t_1/t_1 \log t_2$.

\begin{figure}[pbth]
\centering
\includegraphics[scale=0.5]{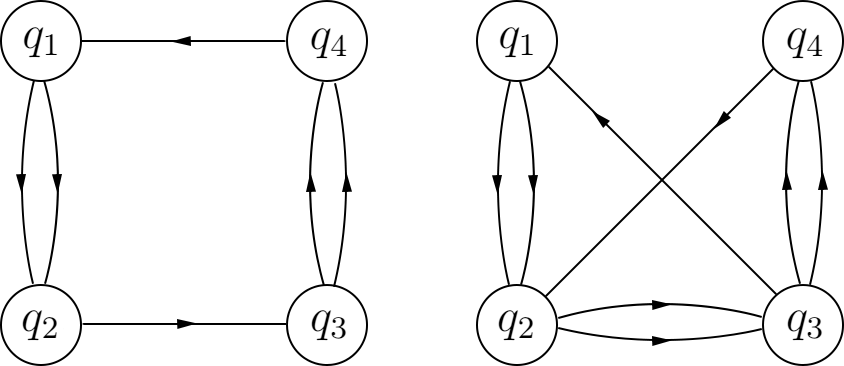}
\caption{\small These two types of quiver diagrams arise from cutting off a pair of
5d BPS states in $SU(3)_{k}$ BPS quiver. The left one is the usual 4d BPS quiver for
4d $SU(3)$ theory. The one on the right is equivalent to the left via quiver
mutation, to be performed on node 4, followed by a switch of the resulting nodes 3 and 4.
This is of course consistent with the fact that locally
the 4d holonomy saddle  is always  the ordinary 4d $SU(3)$ theory.}
\label{Fig-4dSU3-quiver}
\end{figure}

It is sufficient to construct the 4d quiver diagram using the massless states at two of those three singularities and we will choose $P_1$ and $P_2$ in the following. If we approach $P_1$ and $P_2$ from $\mathcal{O}$ we will find the following contours collapse
\begin{itemize}
\item At $P_1$: $B_{12}$ and $B_{23}$
\item At $P_2$: $B_{12}-A_1 + A_2$ and $B_{23}-A_2 + A_3$
\end{itemize}
The local 4d quiver diagram consists four nodes $q_1,\cdots,q_4$ and we may choose the four contours collapsing at $P_1$ and $P_2$ as representatives, namely $B_{12},-B_{12}+ A_1 - A_2,B_{23}$ and $-B_{23}+ A_2 - A_3$.
And the 4d quiver diagram is drawn as the left one in Fig.~\ref{Fig-4dSU3-quiver}.

We need to determine the other two nodes in the 5d quiver diagram and the method is the same as in the previous cases: we consider another holonomy saddle $\mathcal{O}'$ whose local geometry is indistinguishable from the original one at $\mathcal{O}$. Then we may similarly write down the four contours parallel with those at $\mathcal{O}$ and pull them back from $\mathcal{O}'$ to $\mathcal{O}$. 

Consider applying an $\mathbb{Z}_3$ transformation given above to arrive at another holonomy saddle located at $U_1 = 3 e^{\frac{2\pi i}{3}},U_2 = 3 e^{-\frac{2\pi i}{3}}$, as shown in the second diagram in Fig.~\ref{Fig-5dSU30-U-plane}. Note that the two diagrams in Fig.~\ref{Fig-5dSU30-U-plane} represent different hyperplanes in the moduli space parametrized by $U_1,U_2$: the first is $U_1=U_2$ and the second is $U_1 = e^{\frac{4\pi i}{3}} U_2$. Nevertheless, we will call both of them $U$-plane where $U$ is related to $U_1,U_2$ as shown in the figure. The branch cuts and 1-cycles on the spectral curve are depicted in the second diagram in Fig.~\ref{Fig-5dSU30-curve-cycle} and we may consider the local quivers whose four nodes are represented by the contours $B'_{12}, -B'_{12} + A'_1 - A'_2 - [\textrm{KK}], B'_{23}$ and $-B'_{23} + A'_2 - A'_3$, where the KK-charge will be explained soon.

\begin{figure}[pbth]
\centering
\includegraphics[scale=0.35]{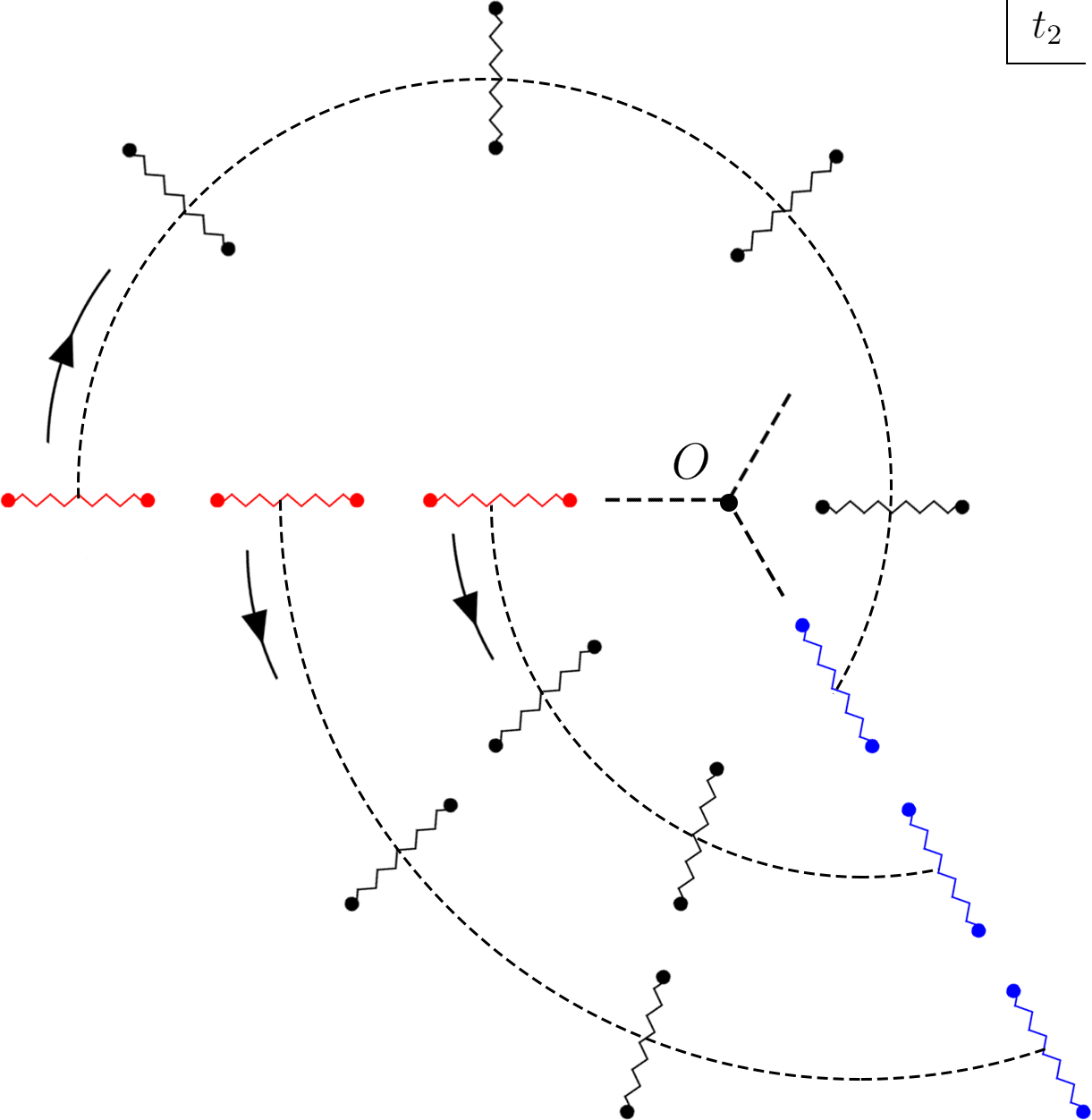}
\caption{\small The transformations of branch cuts on the $t_2$-plane following the path from $\mathcal{O}$ to $\mathcal{O}'$ on the $U$-plane for $SU(3)_0$ theory.}
\label{Fig-5dSU30-curve-cycle-rotation}
\end{figure}

We choose a path from $\mathcal{O}$ to $\mathcal{O}'$ as shown in Fig.~\ref{Fig-5dSU30-U-plane}. Starting at $\mathcal{O}$ we follow the path in the $U_1=U_2$ plane given by the first diagram and move to the origin, then we enter the $U_1 = e^{\frac{4\pi i}{3}} U_2$ plane given by the second diagram and move to the other saddle $\mathcal{O}'$. The branch cuts on the spectral curve will rotate according to Fig.~\ref{Fig-5dSU30-curve-cycle-rotation}.
Initially for $\mathcal{O}$ the three branch cuts are colored in red and they will rotate following the dashed line to reach the blue branch cuts, which corresponds to $\mathcal{O}'$ at the other saddle. The three branch cuts will permute after the rotation, which gives
\begin{equation}
	A_1 \rightarrow A'_2,\quad A_2 \rightarrow A'_3,\quad A_3 \rightarrow A'_1.
\end{equation}
Recall the Seiberg-Witten differential $\lambda_{\textrm{SW}}$ is chosen as $\lambda_{\textrm{SW}} = -(4\pi^2 R_5 i)^{-1} d t_1/t_1 \log t_2$ and we can shift that by a total derivative to make sure it gives the standard 4d $\lambda_{\textrm{SW}} = z dt/(2\pi i t) $ at the first holonomy saddle. Then from Fig.~\ref{Fig-5dSU30-curve-cycle-rotation} we can see that the contribution of $A'_2$ and $A'_3$ will get an additional piece $-i/3R_5$ from the phase of $\log t_2$ while $A'_1$ will get $2i/3R_5$ instead. Therefore the cycle $A'_1 - A'_2$ will carry a KK-charge and that is the reason for the $-[\textrm{KK}]$ term mentioned above.

For $B'$-cycles one finds
\begin{equation}
	B_{12} \rightarrow B'_{23},\quad B_{23} \rightarrow B'_{31} + 2\mathcal{C}_O ,\quad B_{31} \rightarrow B'_{12} - 2\mathcal{C}_O,
\end{equation}
where $\mathcal{C}_O$ represents the cycle enclosing the origin contour-clockwisely which contributes to half the instanton charge. Here we have used the fact that the combination of three $A$-cycles is trivial due to the cancellation of residues at $t_2=0$ and $t_2 = \infty$. Therefore the four contours at $\mathcal{O}'$ will transform backward as
\begin{equation}
	B'_{12} \rightarrow B_{31} + 2 \mathcal{C}_O,\quad -B'_{12} + A'_1 - A'_2 -[\textrm{KK}] \rightarrow -B_{31} + A_3 - A_1 - 2\mathcal{C}_O -[\textrm{KK}]
\end{equation}
and
\begin{equation}
	B'_{23} \rightarrow B_{12},\quad -B'_{23} + A'_2 - A'_3 \rightarrow -B_{12} + A_1 - A_2.
\end{equation}
Note that the pullback of the second pair $B'_{23}$ and $-B'_{23} + A'_2 - A'_3$ are already part of the 4d quivers for $\mathcal{O}$. Therefore we can associate the remaining two nodes $q_5,q_6$ in the 5d quiver diagram with the contours $B_{31} + 2 \mathcal{C}_O$ and $-B_{31} + A_3 - A_1 - 2\mathcal{C}_O -[\textrm{KK}]$.

With these sets of states, the 5d quiver diagram is obtained as the first diagram in Fig.~\ref{Fig-5dSU3-quiver},
and can be seen as an extension of the 4d one in Fig.~\ref{Fig-4dSU3-quiver}. The arrows can be worked out using
the intersection numbers given as,
\begin{align}\label{eq-SU(3)-cycles-intersection}
	A_1 \# B_{12} &= B_{12} \# A_2 = 1,\nonumber \\
	A_2 \# B_{23} &= B_{23} \# A_3 = 1,\nonumber \\
	A_3 \# B_{31} &= B_{31} \# A_1 = 1,
\end{align}
with all others vanishing.

\subsection{$SU(3)_1$}
We then move further to construct the quiver diagram for the $SU(3)_1$ theory based on the previous discussions. The spectral curve is given by \eqref{eq-5dSU21-mirror-curve} as
\begin{equation}
	P_{SU(3)_1} = \sqrt{\lambda} \left(\frac{t_2^2}{t_1} + t_1 t_2^2 \right) + \left(t_2^3 + U_1 t_2^2 + U_2 t_2 + 1 \right) = 0.
\end{equation}
There is no $\mathbb{Z}_3$ symmetry in $SU(3)_1$ theory since the local $\theta$-angles between the two adjacent holonomy saddles differ by $2\pi/3$.

\begin{figure}[pbth]
\centering
\includegraphics[scale=0.6]{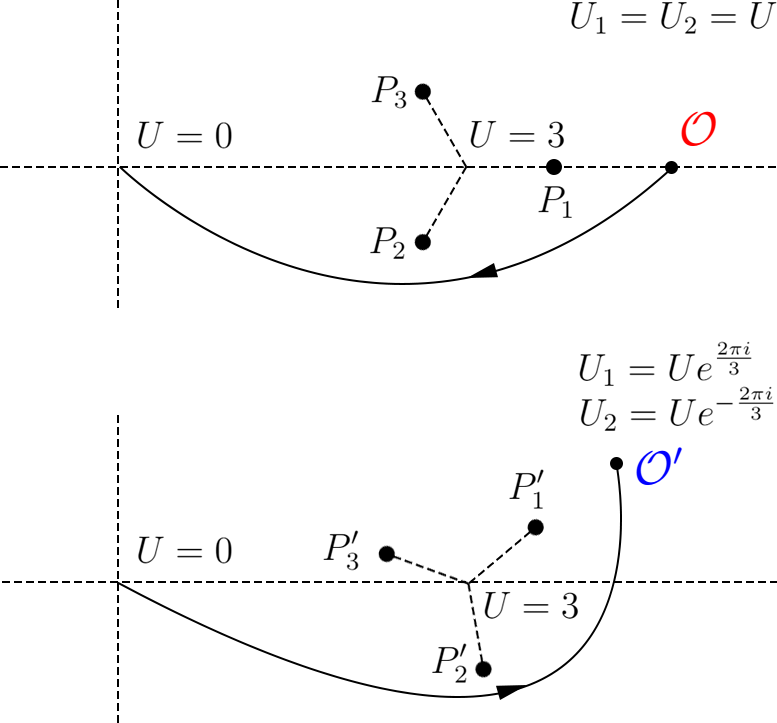}
\caption{\small The $U$-plane for $SU(3)_1$ theory and the path connecting the two observers $\mathcal{O}$ and $\mathcal{O}'$ probing the two holonomy saddles. }
\label{Fig-5dSU31-U-plane}
\end{figure}

The analysis in the following is parallel to $SU(2)_{\pi}$ and $SU(3)_0$ cases. Let's consider the hyperplane $U_1=U_2=U$ which is labeled by $U$. One of the holonomy saddle $\mathcal{O}$ is sitting around $U=3$, and if we consider the 4d limit $|\lambda| \ll 1$ (Equivalently, one fix the QCD scale $\Lambda$ and send $R_5$ to zero) there will be three singularities $P_1,P_2,P_3$ surrounding $U=3$. One can always adjust the phase of $\lambda$ to let one of the singularities sit at the real axis such that the $U$-plane will look the same as that in $SU(3)_0$ theory, which is also depicted as the first diagram in Fig.~\ref{Fig-5dSU31-U-plane}.

\begin{figure}[pbth]
\centering
\includegraphics[scale=0.35]{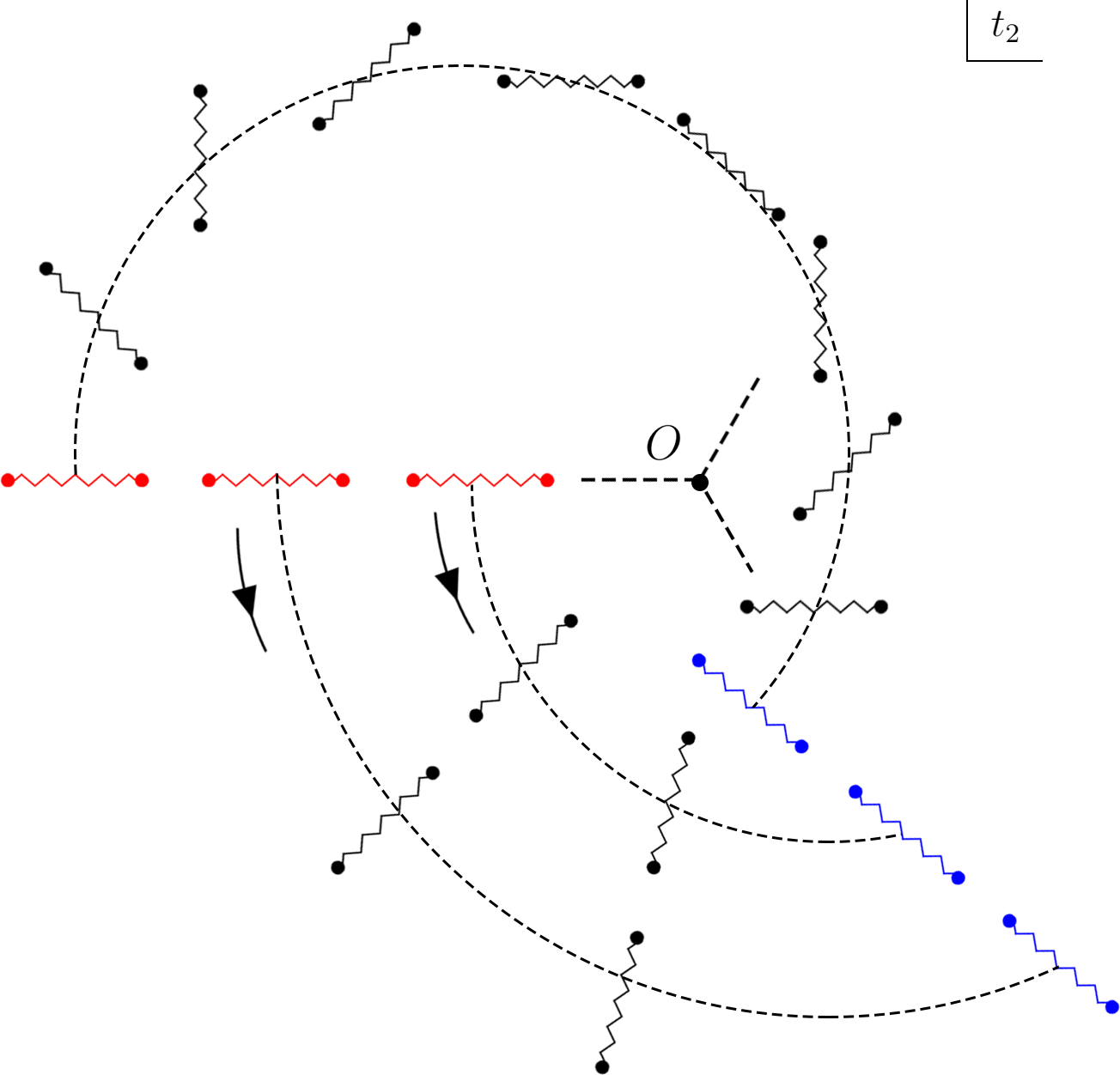}
\caption{\small The transformations of branch cuts on the $t_2$-plane following the path from $\mathcal{O}_B$ to $\mathcal{O}_A$ on the $U$-plane for $SU(3)_1$ theory.}
\label{Fig-5dSU31-curve-cycle-rotation}
\end{figure}

The same complication as in the $SU(2)_{\pi}$ theory arises if we move to the next saddle $\mathcal{O}'$ in the hyperplane $U_1 = e^{\frac{4\pi i}{3}} U_2$ where the local $\theta$-angle is rotated by 120-degree. Following the same discussion below \eqref{eq-5dSU21-theta-angle-redefine} the three singularities $P_1,P_2,P_3$ will rotate a 40-degree (counter-clockwise) which gives the second diagram in Fig.~\ref{Fig-5dSU31-U-plane}. Moreover, the branch cuts on the spectral curve will rotate a 20-degree (counter-clockwise) which gives the blue branch cuts in Fig.~\ref{Fig-5dSU31-curve-cycle-rotation}.

Following the path described in Fig.~\ref{Fig-5dSU31-U-plane} to move from one saddle $\mathcal{O}$ to its counterpart $\mathcal{O}'$, the rotation of branch cuts are sketched in Fig.~\ref{Fig-5dSU31-curve-cycle-rotation}. Initially for $\mathcal{O}$ the three branch cuts are colored in red and they will rotate following the dashed line to reach the blue cuts, which corresponds to $\mathcal{O}'$ at the other saddle.
The choices of $A/B$-cycles are still given in Fig.~\ref{Fig-5dSU30-curve-cycle} except the three cuts in the second diagram need to rotate 20-degree counter-clockwise to match those in Fig.~\ref{Fig-5dSU31-curve-cycle-rotation}. One can work out the transformations of the $A/B$-cycles and we have
\begin{equation}
	A_1 \rightarrow A'_2,\quad A_2 \rightarrow A'_3,\quad A_3 \rightarrow A'_1,
\end{equation}
for $A$-cycles and
\begin{equation}
	B_{12} \rightarrow B'_{23},\quad B_{23} \rightarrow B'_{31} +A'_1 + 2\mathcal{C}_O,\quad B_{31} \rightarrow B'_{12} - A'_1 - 2\mathcal{C}_O,
\end{equation}
for $B$-cycles.

We again choose the contours $B_{12},-B_{12}+A_1-A_2,B_{23}$ and $-B_{23}+A_2-A_3$ as the representatives of the local 4d quiver diagram for the first saddle $\mathcal{O}$, and the 'prime' version $B'_{12},-B'_{12}+A_1-A_2 - [\textrm{KK}] ,B'_{23}$ and $-B'_{23}+A_2-A_3$ for the second saddle $\mathcal{O}'$. Pulling the four contours at $\mathcal{O}'$ backward according to the above transformation we will find
\begin{equation}
	B'_{12} \rightarrow B_{31}+A_3+2\mathcal{C}_O,\quad -B'_{12}+A'_1 - A'_2  - [\textrm{KK}] \rightarrow -B_{31}  -A_1 - 2\mathcal{C}_O  - [\textrm{KK}] ,
\end{equation}
and
\begin{equation}
	B'_{23} \rightarrow B_{12},\quad -B'_{23} + A'_2 - A'_3 \rightarrow -B_{12} + A_1 - A_2.
\end{equation}
The pullback of the second pair $B'_{23}$ and $-B'_{23} + A'_2 - A'_3$ are again part of the 4d quivers for $\mathcal{O}$. Therefore we can associate the six nodes $q_1,\cdots,q_6$ in the 5d BPS quiver diagram with
the contours $B_{12}$, $-B_{12}+A_1-A_2$, $B_{23}$, $-B_{23}+A_2-A_3$, $B_{31}+A_3+2\mathcal{C}_O$, and $-B_{31}  -A_1 - 2\mathcal{C}_O  - [\textrm{KK}]$.
The 5d quiver diagram is then drawn as the top-right diagram in Fig.~\ref{Fig-5dSU3-quiver} with
the intersection numbers found again via \eqref{eq-SU(3)-cycles-intersection}.

\subsection{$SU(3)_2$ and $SU(3)_3$}

In the last part of this section, we will briefly show the results of $SU(3)_2$ and $SU(3)_3$ for completeness. The spectral curves are
\begin{equation}
	P_{SU(3)_2} = \sqrt{\lambda} \left(\frac{t_2^2}{t_1} + t_1 t_2^3 \right) + \left(t_2^3 + U_1 t_2^2 + U_2 t_2 + 1 \right) = 0,
\end{equation}
for $SU(3)_2$ and
\begin{equation}
	P_{SU(3)_3} = \sqrt{\lambda} \left(\frac{t_2^3}{t_1} + t_1 t_2^3 \right) + \left(t_2^3 + U_1 t_2^2 + U_2 t_2 + 1 \right) = 0,
\end{equation}
for $SU(3)_3$\footnote{We consider $|\lambda| \ll 1$ such that $f_3(\lambda)\approx \sqrt{\lambda}$ in \eqref{eq-f-lambda}.}. For each theory there are three holonomy saddles on the moduli space where the local theories are 4d $SU(3)$ theory. For $SU(3)_2$ the adjacent two saddles are glued with a relative shift of 4d $\theta$-angle by $4\pi/3$ and for $SU(3)_3$ theory by $2\pi$. Therefore the $\mathbb{Z}_3$ symmetry is preserved in $SU(3)_3$ theory but is broken in $SU(3)_2$ theory.

\begin{figure}[pbth]
\centering
\includegraphics[scale=0.6]{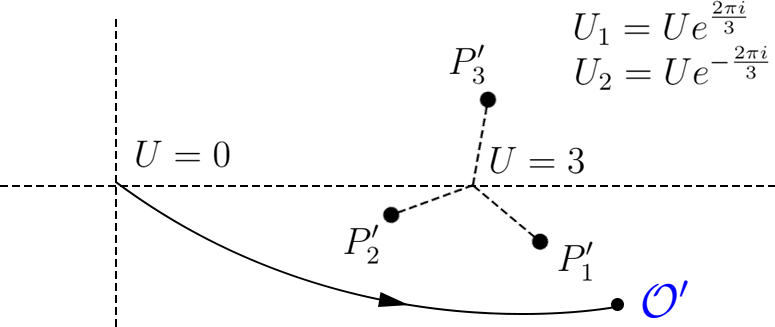}
\caption{\small The $U$-plane for $SU(3)_2$ theory and the path connecting the two observers $\mathcal{O}$ and $\mathcal{O}'$ probing the two holonomy saddles. }
\label{Fig-5dSU32-U-plane}
\end{figure}

We choose the $U$-plane in the same way as before and in the 4d limit $|\lambda| \ll 1$ there are three singularities on the $U$-plane at each holonomy saddles. We will adjust the phase of $\lambda$ such that at the first saddle one of the singularities sits at the real axis just as in the previous cases, for example, as shown in the first diagram in Fig.~\ref{Fig-5dSU20-U-plane-path} and \ref{Fig-5dSU21-U-plane}. Starting at $\mathcal{O}$, we will follow a path through the origin in the moduli space to move to its counterpart at the second holonomy saddle. For $SU(3)_3$ the path is the same to the $SU(3)_0$ theory shown in Fig.~\ref{Fig-5dSU20-U-plane-path} and for $SU(3)_2$ the destination $\mathcal{O}'$ is shown in Fig.~\ref{Fig-5dSU32-U-plane}.

The 5d quiver diagrams can be constructed in the same way by combining the two local 4d quivers at the two holonomy saddles. Choosing the $A/B/A'/B'$ cycles in a similar way, one finds for both $SU(3)_2$ and $SU(3)_3$ the $A$-cycles just permute as before
\begin{equation}
	A_1 \rightarrow A'_2,\quad A_2 \rightarrow A'_3,\quad A_3 \rightarrow A'_1.
\end{equation}
For $SU(3)_2$ theory the $B$-cycles transform as
\begin{equation}
	B_{12} \rightarrow B'_{23} + A'_2 - A'_3,\quad B_{23} \rightarrow B'_{31} -A'_2 + 2\mathcal{C}_O,\quad B_{31} \rightarrow B'_{12} + A'_3 - 2\mathcal{C}_O,
\end{equation}
and for $SU(3)_3$ theory the $B$-cycles transform as
\begin{equation}
	B_{12} \rightarrow B'_{23} + A'_2 - A'_3,\quad B_{23} \rightarrow B'_{31} + A'_1 - A'_2 + 2\mathcal{C}_O,\quad B_{31} \rightarrow B'_{12} - A'_1 + A'_3 - 2\mathcal{C}_O,
\end{equation}
where we have used the fact that $A_1+A_2+A_3$ (or $A'_1+A'_2+A'_3$) is trivial. In the present cases we will associate the four nodes $q_1,\cdots,q_4$ in the 5d quiver diagram with the contours $B_{12}, -B_{12}+ A_1 - A_2, B_{23}+ A_2 - A_3$ and $-B_{23}$ for both theories. The 4d quiver diagram is depicted as the right one in Fig.~\ref{Fig-4dSU3-quiver} which is related to the left one via a quiver mutation. At the other saddle $\mathcal{O}'$ the four contours are just the 'prime' version $B'_{12}, -B'_{12}+ A'_1 - A'_2 - [\textrm{KK}], B'_{23}+ A'_2 - A'_3$ and $-B'_{23}$, and if we pull the four contours backward following the path described in Fig.~\ref{Fig-5dSU20-U-plane-path} or \ref{Fig-5dSU32-U-plane} we will find the following results. For $SU(3)_2$ theory they will become
\begin{equation}
	B'_{12} \rightarrow B_{31}-A_2 + 2\mathcal{C}_O, \quad -B'_{12}+A'_1 - A'_2 - [\textrm{KK}] \rightarrow -B_{31} - 2 A_1 - 2\mathcal{C}_O - [\textrm{KK}],
\end{equation}
and
\begin{equation}
	B'_{23} + A'_2 - A'_3 \rightarrow B_{12}, \quad -B'_{23} \rightarrow -B_{12} + A_1 - A_2.
\end{equation}
The pullback of the second pair $B'_{23}+A'_2-A'_3$ and  $-B'_{23}$ are part of the 4d quivers for $\mathcal{O}$. Therefore the remaining two nodes $q_5,q_6$ in the 5d quiver diagram for $SU(3)_2$ theory are associated with the contours $B_{31}-A_2 + 2\mathcal{C}_O$ and $-B_{31} - 2 A_1 - 2\mathcal{C}_O - [\textrm{KK}]$.
The 5d quiver diagram is depicted as the bottom-left diagram in Fig.~\ref{Fig-5dSU3-quiver}.

For $SU(3)_3$ theory the $B$-cycles transform as
\begin{equation}
	B'_{12} \rightarrow B_{31}-A_2 + A_3 + 2\mathcal{C}_O, \quad -B'_{12}+A'_1 - A'_2  - [\textrm{KK}] \rightarrow -B_{31} - A_1 + A_2 - 2\mathcal{C}_O  - [\textrm{KK}],
\end{equation}
and
\begin{equation}
	B'_{23} + A'_2 - A'_3 \rightarrow B_{12}, \quad -B'_{23} \rightarrow -B_{12} + A_1 - A_2.
\end{equation}
The pullback of the second pair $-B'_{23}$ and $B'_{23}+A'_2-A'_3$ are again part of the 4d quivers for $\mathcal{O}$ and we can associate the remaining two nodes $q_5,q_6$ in 5d quiver diagram with the contours $B_{31}-A_2 + A_3 + 2\mathcal{C}_O$ and $-B_{31} - A_1 + A_2 - 2\mathcal{C}_O  - [\textrm{KK}]$.
The 5d quiver diagram is depicted as the bottom-right diagram in Fig.~\ref{Fig-5dSU3-quiver} where the $\mathbb{Z}_3$ symmetry is also manifest.

\section{Duality of $SU(N)_{N}$}

In the preceding sections, we constructed 5d BPS quivers by noting how
each 4d holonomy saddle supplies a 4d BPS quiver, and how the relation between
adjacent 4d saddles supplies two additional nodes which are needed to complete
the 5d quiver. In all of these procedures, the two additional 5d nodes would
be labeled as the $(2n-1)$-th and $2n$-th for $n=1,2,\cdots, N$ which itself
labels the $N$ 4d holonomy saddles. Although this by itself is universal for
all level, $SU(N)_0$ is special in that no matter which $n$ we take, the excised
5d BPS quiver produces the same 4d BPS quiver, rather than modulo mutation.
The spectral curve for $SU(N)_0$ also admits $\IZ_N$ symmetry manifestly
where $U$'s shift between 4d saddles, naturally.

For $SU(N)_N$ with $\IZ_{{\rm gcd}(N,N)}=\IZ_N$, on the other hand, the shape
of the 5d BPS quivers we have found are clearly more symmetric beyond this
$\IZ_N$.  To see the origin of this phenomenon, it is
more instructive to generalize an alternate
form of the F0 curve, rather than its equivalent F2 curve, to higher rank. The
F0 curve we used was
\bea
\sqrt{\lambda}\,(\tilde t_2\tilde t_1+\tilde t_2/\tilde t_1)+(\tilde t_2^{\,2}+U\tilde t_2+1)=0
\eea
where we introduced a tilded $t$'s to avoid a confusion with $t$ variables for
the $SU(N)_N$ curve below. It may be more suggestively written, after division by $\tilde t_2\lambda^{1/4}$,
\bea
U\lambda^{-1/4}+(\lambda^{1/4}\tilde t_1+ \lambda^{-1/4}\, \tilde t_2)+(\lambda^{1/4}\, \tilde t_1^{-1}+\lambda^{-1/4} \tilde t_2^{-1} )=0
\eea
With $V\equiv \lambda^{-1/4}U$, $s_1\equiv \lambda^{1/4}\tilde t_1$,
and $s_2\equiv \lambda^{-1/4}\tilde t_2$, this
gives
\bea
V+(s_1+s_2)+ \frac{\lambda^{1/2}}{s_1}+\frac{1}{\lambda^{1/2} s_2}=0
\eea
This form of F0 curve is invariant under
\bea
B:\qquad s_{1,2}\rightarrow s_{2,1}\ , \quad \lambda^{1/2}\rightarrow \lambda^{-1/2}
\eea
This changes the coupling $\lambda$, so it is not  a symmetry of the
field theory, strictly speaking; it would be more appropriately called
a duality.

In the $(p,q)$ 5-brane web realization, the map $B$ can be seen as
a swap of the vertical direction and the horizontal direction, or the swap of D5-branes and NS5-branes.
The invariant value $\lambda=1$ would correspond to
a square form of the internal face, at which the $W$-boson represented
by the F-string segment, and a charged instanton represented by the D-string segment, would become
massless simultaneously. With $\lambda> 1$, the $SU(2)$ symmetry restoration
occurs with massless $W$-boson, and $B$ maps this to $\lambda < 1$ where a charged
instanton plays the role of $W$-boson instead. This well-known duality extends
all the way to the strict 5d limit of $R_5\rightarrow\infty$, where $\lambda$
is real and positive.

For higher ranks, the usual $(p,q)$ brane-webs thereof do not show such a duality
manifestly. On the other hand, we may try to map the generalization of F2 curve
\begin{equation}
\frac{1}{\lambda^{1/2}+\lambda^{-1/2}}\left(t_1 + t_1^{-1} \right) + \left(t_2^N + U_1 t_2^{N-1} + U_2 t_2^{N-2} + \cdots +U_{N-1} t_2 + 1  \right)=0,
\end{equation}
to a form similar to the above alternate version of F0, with the new coordinates
$s$'s and the Coulomb vev $V$'s as
\begin{equation}
	t_1 = \frac{s_1}{\lambda^{1/2} s_2},\quad t_2 = \frac{s_1+s_2}{(\lambda^{1/2}+\lambda^{-1/2})^{\frac{1}{N}}}\ , \quad U_a = \frac{V_a}{(\lambda^{1/2}+\lambda^{-1/2})^{\frac{a}{N}}}\ .
\end{equation}
The curves is now rewritten as
\begin{equation}
V_{N-1} + V_{N-2}(s_1+s_2) + \cdots + V_1 (s_1+s_2)^{N-2}+(s_1+s_2)^{N-1} + \frac{\lambda^{1/2}}{s_1}+ \frac{1}{\lambda^{1/2}s_2}=0,
\end{equation}
with $d \lambda_{\textrm{SW}}$ retaining the same general form both in $t$
coordinates and in $s$ coordinates. The above coordinate transformation is actually the Hanay-Witten transition \cite{Hanany:1996ie} in terms of the 5d brane web.

The expected $\IZ_{N= {\rm gcd}(N,N)}$ symmetry manifest here no differently
as\footnote{At $\lambda=-1$, this $\IZ_N$ symmetry extends to a $\IZ_{2N}$
symmetry, as can be seen easily in the latter variable choice,
\bea
C:\qquad s_{1,2}\rightarrow  s_{2,1}\, e^{\pi i/N}\ , \quad V_a\rightarrow V_a \, e^{a\pi i/N}
\eea
This manifests in the singularity distribution in the Coulomb moduli space. See
the Appendix for a related comment at the end.}
\bea
A:\qquad s_{1,2}\rightarrow s_{1,2}\,e^{2\pi i/N}\ , \quad
V_a \rightarrow V_a \,e^{2\pi a i/N}
\eea
while, allowing $\lambda$ to transform as well, we find again a $\IZ_2$ duality
\bea
B:\qquad s_{1,2}\rightarrow s_{2,1} \ ,\quad \lambda^{1/2}\rightarrow \lambda^{-1/2}
\eea
The latter $\IZ_2$ generated by $B$
flips $\lambda^{1/2}\rightarrow \lambda^{-1/2}$, so
$\lambda = 1$ is a self-dual point in that the transformation
maps within the same Seiberg-Witten moduli space. Otherwise, the map
relates theories with $|\lambda|>1$ to those with $|\lambda|<1$. The
quantity $\log(\lambda)$ is proportional to the bare inverse coupling
squared in the 5d sense, so the map relates Seiberg-Witten theory with
positive and negative values of 5d bare coupling squared. The latter does not lead
to inconsistency, as is well known, by the Coulomb phase being cut off
before the effective coupling turns negative.

$\IZ_N$ generated by $A$ acts with $\lambda$ fixed, so may be considered as a symmetry
that relates different Coulombic vacua and thus is capable of moving between holonomy
saddles. The above $\IZ_2$ generally changes $\lambda$, so cannot be a map
among 4d holonomy saddles of a single Seiberg-Witten theory.
What does happen with $B$ is that, across $|\lambda|=1$, the
nature of $N$ 4d holonomy saddles changes. In the strict 5d limit, this connects
with how $SU(2)$ symmetry restoration happens differently depending on the sign
of $\log(|\lambda|)$; the light charged vector meson that  becomes massless at
the symmetric restoration point is either a fundamental $W$-boson or a
charged instanton, depending on the sign of $\log(|\lambda|)$. $B$ swaps these
two regimes to each other.

Once we move to  $|\lambda|>1$ region, i,e. to the region where the inverse bare 
coupling squared $\mu_0$ becomes negative, the charge states that become light
in the 4d holonomy saddles are different from $|\lambda|<1 $ region. In the simplest examples of F0,
either the 2nd and the 3rd nodes of the 5d BPS quiver or the 4th and the 1st
would remain light in the infrared if we insist on $|\lambda|\gg 1$ \cite{Closset:2021lhd}.

This suggests that, for general $SU(N)_N$, BPS states which become light
in the 4d saddles are $2i$-th and $(2i+1)$-th nodes, for
integer $i$ mod $N$ except for one pair, instead of $(2i-1)$-th and $2i$-th pairs
which we have seen in the previous two sections. 
Although
we still have $N$ holonomy saddles for $|\lambda|\gg 1$, the 4d BPS quiver at the
4d holonomy saddles would be obtained by dropping from 5d BPS quiver
$2n$-th and $(2n+1)$-th nodes for some $n$ mod $N$ instead of $(2n-1)$-th and
$2n$-th nodes. Again, the mutation plays an important role in bringing
us back to the standard 4d $SU(N)$ BPS quivers.

\section*{Acknowledgments}

We would like to thank Cyril Closset for drawing our attention to Ref.~\cite{Closset:2021lhd}
and also for related discussions. QJ and PY were supported by KIAS Individual Grants (PG080801 and PG005704)
at Korea Institute for Advanced Study.

\appendix

\section{An Alternate View on F0-theory}

Ref.~\cite{Closset:2021lhd} analyzed the F0-theory by solving the Picard-Fuchs equation with some special values of $\lambda$ and performing computations that largely relied on the monodromies around singularities on the U-plane.
Here we assume $0<\lambda<1$ more generally and briefly repeat their analysis and give some more explicit results.
In particular, we aim at illustrating how our choice of the basis leads to the quiver Fig.~\ref{Fig-5dSU2-quiver-all}
is found in this alternate view.

The Picard-Fuchs equation of the F0 spectral curve \eqref{eq-5dSU20-mirror-curve} is given by
\begin{equation}
	\frac{\partial^2 \Omega}{\partial w^2} + \frac{4 (8 w-A-B)}{(4 w-A) (4 w-B)} \frac{\partial \Omega}{\partial w} + \frac{4}{(4 w-A) (4 w-B)} \Omega = 0,
\end{equation}
where one works with $w = U^2/16$ due to the $\mathbb{Z}_2$ symmetry $U\rightarrow -U$. $\Omega$ denotes the period of the spectral curve and the parameters $A$ and $B$ are given by
\begin{equation}
	A = (1+\lambda) + 2\sqrt{\lambda},\quad B=(1+\lambda) - 2 \sqrt{\lambda}.
\end{equation}
The magnetic and electric central charges $a_D$ and $a$ are related to the periods as
\begin{equation}\label{eq-5dSU20-periods}
	\Omega_{a_D} = \frac{d a_D}{d U} = \frac{\sqrt{w}}{2}\frac{ d a_D}{ d w},\quad \Omega_a = \frac{\sqrt{w}}{2} \frac{d a}{d w}.
\end{equation}
Solving the Picard-Fuchs equation one obtains the solutions of periods on the $w$-plane
\begin{equation}\label{eq-5dSU20-periods-aD}
	 \frac{d a_D}{d w} = \frac{i}{2\pi R_5} \frac{1}{\sqrt{w(w-\frac{(1-\sqrt{\lambda})^2}{4})}}\   {}_1F_2 \left(\frac{1}{2},\frac{1}{2},1;\frac{4w - (1+\sqrt{\lambda})^2}{4w-(1-\sqrt{\lambda})^2} \right),
\end{equation}
and
\begin{equation}\label{eq-5dSU20-periods-a}
	 \frac{d a}{d w} = \frac{1}{4\pi R_5 } \frac{1}{\sqrt{w(w-\frac{(1-\sqrt{\lambda})^2}{4})}}\   {}_1F_2 \left(\frac{1}{2},\frac{1}{2},1;\frac{4\sqrt{\lambda}}{4w - (1-\sqrt{\lambda})^2} \right),
\end{equation}
where the coefficients can be fixed by their values in the asymptotic region on
the moduli space. From these one can solve $a$ and $a_D$ on the $w$-plane.

\begin{figure}[pbth]
\centering
\includegraphics[scale=0.5]{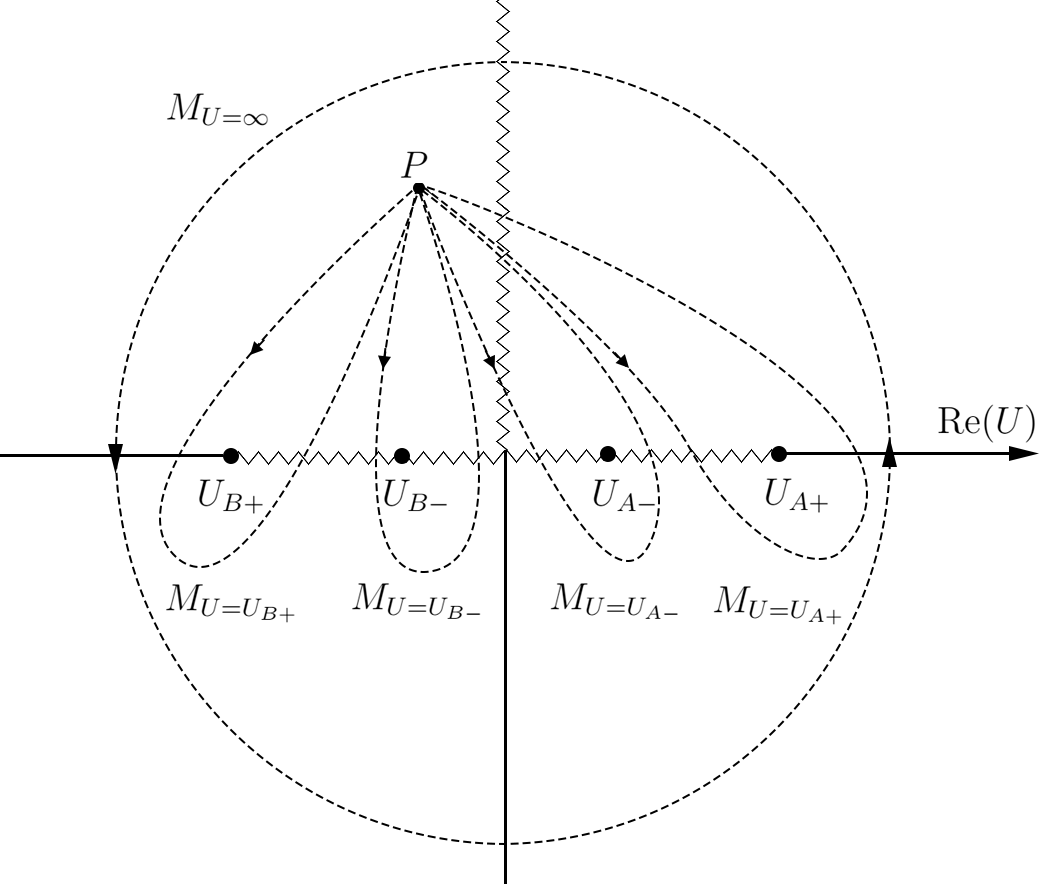}
\caption{\small The contours and monodromies on U-plane for $SU(2)_0$ theory.}
\label{Fig-5dSU20-Monodromy-U-2}
\end{figure}

The full $U$-plane can be obtained by gluing two $w$-plane, the left half-plane and right half-plane in Fig.~\ref{Fig-5dSU20-Monodromy-U-2}, along the imaginary axis in the lower plane such that $a$ and $a_D$ vary smoothly. The branch cuts are inherited from the $w$-plane and are drawn in a symmetric way. Although the origin $U=0$ is a branch point, the monodromy around $U=0$ is actually trivial. There are four singularities $U = U_{A\pm},U_{B\pm}$ on the $U$-plane and we choose the base point $P$ in the second quadrant. Denoting the magnetic and the electric central charges as a column vector
$(a_D,a)^T$, and the $2\times 2$ monodromy matrix $M$ acting the pair
can be read off immediately from the above explicit solution. Since
the above are actually $w$ derivatives, we need to work a little more
to include the instanton and KK-charges in the monodromy matrix.

We can complete the full $4\times 4$ monodromy including instanton and KK-charge. The full central charges are collected as a column vector 	$(a_D,a,i/R_5,\mu_0)^T$, where $i/R_5$ and $\mu_0$ are the central charges of a unit KK mode and a bare instanton. Going around a singularity on the $U$-plane, the central charges are shifted as $M \cdot (a_D,a,i/R_5,\mu_0)^T$ where $M$ is the extended $4 \times 4$ monodromy matrix associated to the singularity. First, notice that moving along a large circle in the asymptotic region of $U$-plane amounts to shifting the electric central charge by $a \rightarrow a+\frac{i}{R_5}$. Since the magnetic central charge $a_D$ is approximated as $i R_5 2a (2a + \mu_0)$, one can read the monodromy $M_{\infty}$ as
\begin{equation}
M_{\infty}=\left(
\begin{array}{cccc}
1&-8&-4&-2\\
0&1&1&0\\
0&0&1&0\\
0&0&0&1
\end{array}
\right),
\end{equation}
We assume the two singularities at $U=U_{B\pm}$ correspond to purely 4d monopole and dyon points, namely one has
\begin{equation}
M_{U_{B+}}=\left(
\begin{array}{cccc}
1&0&0&0\\
-1&1&0&0\\
0&0&1&0\\
0&0&0&1
\end{array}
\right),\quad
M_{U_{B-}}=\left(
\begin{array}{cccc}
-1&4&0&0\\
-1&3&0&0\\
0&0&1&0\\
0&0&0&1
\end{array}
\right).
\end{equation}
Using the fact that $U_{A\pm}$ are conjugated to $U_{B\pm}$ in terms of $w$-plane, one can also work out $M_{U_{A\pm}}$ using $M_{\infty}$ and $M_{U_{B\pm}}$
\begin{equation}
M_{U_{A+}}=\left(
\begin{array}{cccc}
-3&16&4&4\\
-1&5&1&1\\
0&0&1&0\\
0&0&0&1
\end{array}
\right),\quad
M_{U_{A-}}=\left(
\begin{array}{cccc}
-1&4&0&2\\
-1&3&0&1\\
0&0&1&0\\
0&0&0&1
\end{array}
\right).
\end{equation}

\begin{figure}[pbth]
\centering
\includegraphics[scale=0.6]{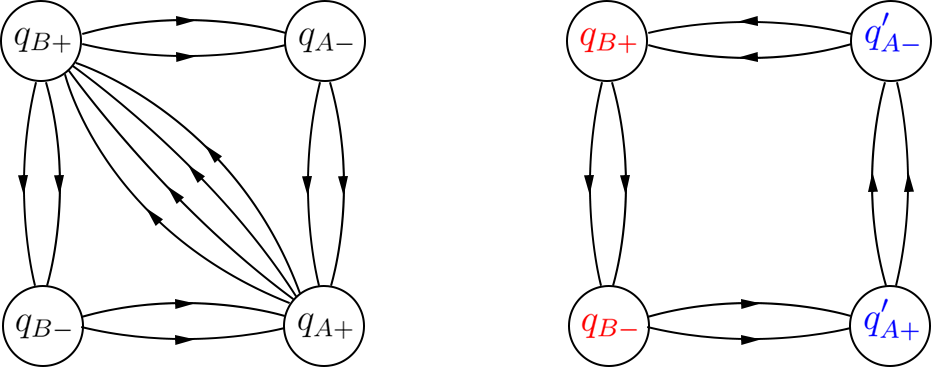}
\caption{\small The quiver diagrams for 5d $SU(2)_0$ theory. The left diagram is constructed according to the above analysis and the right diagram is obtained via a quiver mutation of $q_{A-}$ node. }
\label{Fig-5dSU20-quiver}
\end{figure}

The four singularities are due to massless BPS particles whose charges are inert
under the respective monodromy, which can be read off as
\begin{itemize}
\item $U = U_{B+}$\quad : \quad A monopole of charge $q_{B+}=(1,0,0,0)$, becomes massless
\item $U = U_{B-}$\quad : \quad A dyon of charge $q_{B-}=(-1,2,0,0)$, becomes massless
\item $U = U_{A-}$\quad : \quad A dyon of charge $q_{A-}=(-1,2,0,1)$, becomes massless
\item $U = U_{A+}$\quad : \quad A dyon of charge $q_{A+}=(1,-4,-1,-1)$, becomes massless
\end{itemize}
If we take these as the basis BPS states for the quiver, the 5d BPS quiver
should look like the left of Fig.~\ref{Fig-5dSU20-quiver} \cite{Closset:2021lhd}.
A natural question is how does this choice of basis differ from those
we found in Sec 3 and the underlying reason.

\begin{figure}[pbth]
\centering
\includegraphics[scale=0.7]{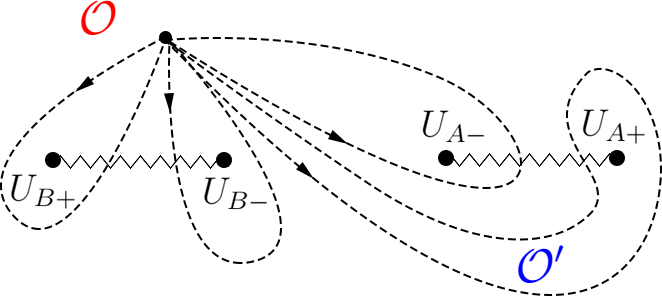}
\caption{\small The monodromies associated with the nodes in the 5d BPS quiver. The 3rd and the 4th paths
become simple when we move from $\mathcal O$ to  $\mathcal O'$ which amounts to the motion from
the first saddle to the second, adopted in Sec. 3 above. These paths differ from those of
Ref.~\cite {Closset:2021lhd}, which explains the difference in the resulting quiver.}
\label{Fig-5dSU20-U-plane-path-OA-OB}
\end{figure}

Note how this quiver is related to the standard one in Fig.~\ref{Fig-5dSU2-quiver-all}
or the one on the right of Fig.~\ref{Fig-5dSU20-quiver} via a quiver mutation of
the $q_{A-}$ node
\begin{equation}
q'_{A-} = -q_{A-}=(1,-2,0,-1),\quad q'_{A+} = q_{A+} + 2 q_{A-} = (-1,0,-1,1).
\end{equation}
What are the corresponding contours on the $U$-plane for $q'_{A-}$ and $q'_{A+}$? Note that in the second quiver diagram the symmetry between two holonomy saddles for F0-theory is manifest and the local 4d quivers are represented by the two blue nodes or the two red nodes. Following the same idea in section 3, let's consider two observers probing the two holonomy saddles on the $U$-plane, sitting at $\mathcal{O}'$ and $\mathcal{O}$ which are symmetric under $U\rightarrow -U$ as shown in Fig.~\ref{Fig-5dSU20-U-plane-path}. Since the quiver diagram is drawn at a specific point on the moduli space, one of the observers, let's say observer $\mathcal{O}'$, must move to the other observer $\mathcal{O}$ and combine their local 4d quivers to form the 5d quiver. That implies the following choices of the last contour enclosing $U_{A+}$ in Fig.~\ref{Fig-5dSU20-U-plane-path-OA-OB} such that if we change the base point to $\mathcal{O}'$ these two contours enclosing $U_{A\pm}$ will become standard but the contour enclosing $U_{B+}$ will become involved instead.

Among the four contours, the two enclosing $U_{B+},U_{B-}$ and $U_{A-}$ are the same as before and their monodromy are still given by $M_{U_{B+}}, M_{U_{B-}}$ and $M_{U_{A-}}$.
The last one is
\begin{equation}
M'_{U_{A+}}  = M_{U_{A-}} \cdot M_{U_{A+}} \cdot M^{-1}_{U_{A-}} = \left( \begin{array}{cccc}
1&0&0&0\\
-1&1&-1&1\\
0&0&1&0\\
0&0&0&1
\end{array}\right),
\end{equation}
and the charge vector $q'_{A+} = (-1,0,-1,1)$ is indeed consistent with this monodromy.

One way to motivate the latter set of contours and the monodromies in the $U$ plane
is to imagine moving from the base point $\cO$ to $\cO'$, through the middle region
between $U_{A_\pm}$ and $U_{B_\pm}$. The somewhat convoluted contours at $\cO$ enclosing
$U_{A_\pm}$ would become simplified as the base point moves to $\cO'$. In fact, if
we further rotate the entire plane 180 degrees so that $\cO'$ moves into the position
of $cO$ and $U_{A_\pm}$ into $U_{B_\pm}$, the newly deformed contour would look
identical to the two original contours  at $\cO$ enclosing $U_{B_\pm}$. $\cO$
and $\cO'$ each probes local geometry of the two 4d holonomy saddles in our language,
so this means that $q'_{A_\pm}$ play exactly the same role in the second saddle as
$q_{B_\pm}$ do in the first saddle. The latter criterion is how we picked up
the 3rd and the 4th BPS states to complete the 5d BPS quiver at $\cO$, so it is
no surprise that we reproduce  Fig.~\ref{Fig-5dSU2-quiver-all} from this alternate
set of contour and monodromies.

There is a special case where $\lambda=-1$ and the $\mathbb{Z}_2$ symmetry of the $U$-plane is enhanced to $\mathbb{Z}_4$, which is studied in detail in \cite{Closset:2021lhd}. When $\lambda=-1$ the discriminant of the spectral curve is reduced to
\begin{equation}
	\Delta(U,\lambda=-1) = U^4 + 64,
\end{equation}
such that the four singularities on the $U$-plane are located at $U = 2\sqrt{2} e^{\frac{(2k+1) \pi i}{4}}$ where $k=0,1,2,3$. The corresponding BPS states will give the $\mathbb{Z}_4$-symmetric quiver as shown in the right of Fig.~\ref{Fig-5dSU20-quiver}.


\begin{thebibliography}{99}



\bibitem{Davies:1999uw}
N.~M.~Davies, T.~J.~Hollowood, V.~V.~Khoze and M.~P.~Mattis,
``Gluino condensate and magnetic monopoles in supersymmetric gluodynamics,''
Nucl. Phys. B \textbf{559}, 123-142 (1999)
[arXiv:hep-th/9905015 [hep-th]].

\bibitem{Lee:1997vp}
K.~M.~Lee and P.~Yi,
``Monopoles and instantons on partially compactified D-branes,''
Phys. Rev. D \textbf{56}, 3711-3717 (1997)
[arXiv:hep-th/9702107 [hep-th]].

\bibitem{Closset:2017bse}
C.~Closset, H.~Kim and B.~Willett,
``$ \mathcal{N} $ = 1 supersymmetric indices and the four-dimensional A-model,''
JHEP \textbf{08}, 090 (2017)
[arXiv:1707.05774 [hep-th]].


\bibitem{Hwang:2018riu}
C.~Hwang, S.~Lee and P.~Yi,
``Holonomy Saddles and Supersymmetry,''
Phys. Rev. D \textbf{97}, no.12, 125013 (2018)
[arXiv:1801.05460 [hep-th]].

\bibitem{Hwang:2017nop}
C.~Hwang and P.~Yi,
``Twisted Partition Functions and $H$-Saddles,''
JHEP \textbf{06}, 045 (2017)
[arXiv:1704.08285 [hep-th]].





\bibitem{Witten:1982df}
E.~Witten,
``Constraints on Supersymmetry Breaking,''
Nucl. Phys. B \textbf{202}, 253 (1982)






\bibitem{Duan:2020qjy}
Z.~Duan, D.~Ghim and P.~Yi,
``5D BPS Quivers and KK Towers,''
JHEP \textbf{02}, 119 (2021)
[arXiv:2011.04661 [hep-th]].



\bibitem{Seiberg:1994rs}
N.~Seiberg and E.~Witten,
``Electric - magnetic duality, monopole condensation, and confinement in N=2 supersymmetric Yang-Mills theory,''
Nucl. Phys. B \textbf{426}, 19-52 (1994)
[erratum: Nucl. Phys. B \textbf{430}, 485-486 (1994)]
[arXiv:hep-th/9407087 [hep-th]].

\bibitem{Seiberg:1994aj}
N.~Seiberg and E.~Witten,
``Monopoles, duality and chiral symmetry breaking in N=2 supersymmetric QCD,''
Nucl. Phys. B \textbf{431}, 484-550 (1994)
[arXiv:hep-th/9408099 [hep-th]].







\bibitem{Ganor:1996pc}
O.~J.~Ganor, D.~R.~Morrison and N.~Seiberg,
``Branes, Calabi-Yau spaces, and toroidal compactification of the N=1 six-dimensional E(8) theory,''
Nucl. Phys. B \textbf{487}, 93-127 (1997)
[arXiv:hep-th/9610251 [hep-th]].





\bibitem{Staudacher:2000gx}
M.~Staudacher,
``Bulk Witten indices and the number of normalizable ground states in supersymmetric quantum mechanics of orthogonal, symplectic and exceptional groups,''
Phys. Lett. B \textbf{488}, 194-198 (2000)
[arXiv:hep-th/0006234 [hep-th]].

\bibitem{Pestun:2002rr}
V.~Pestun,
``N=4 SYM matrix integrals for almost all simple gauge groups (except E(7) and E(8)),''
JHEP \textbf{09}, 012 (2002)
[arXiv:hep-th/0206069 [hep-th]].




\bibitem{Lee:2017lfw}
S.~J.~Lee and P.~Yi,
``D-Particles on Orientifolds and Rational Invariants,''
JHEP \textbf{07}, 046 (2017)
[arXiv:1702.01749 [hep-th]].

\bibitem{Witten:2000nv}
E.~Witten,
``Supersymmetric index in four-dimensional gauge theories,''
Adv. Theor. Math. Phys. \textbf{5}, 841-907 (2002)
[arXiv:hep-th/0006010 [hep-th]].

\bibitem{Aharony:2013hda}
O.~Aharony, N.~Seiberg and Y.~Tachikawa,
``Reading between the lines of four-dimensional gauge theories,''
JHEP \textbf{08}, 115 (2013)
[arXiv:1305.0318 [hep-th]].



\bibitem{Jia:2021ikh}
Q.~Jia and P.~Yi,
``Aspects of 5d Seiberg-Witten theories on $\IS^1$,''
JHEP \textbf{02}, 125 (2022)
[arXiv:2111.09448 [hep-th]].


\bibitem{Closset:2019juk}
C.~Closset and M.~Del Zotto,
``On 5d SCFTs and their BPS quivers. Part I: B-branes and brane tilings,''
[arXiv:1912.13502 [hep-th]].

\bibitem{Closset:2021lhd}
C.~Closset and H.~Magureanu,
``The $U$-plane of rank-one 4d $\mathcal{N}=2$ KK theories,''
[arXiv:2107.03509 [hep-th]].
\bibitem{Witten:1982fp}
E.~Witten,
``An SU(2) Anomaly,''
Phys. Lett. B \textbf{117}, 324-328 (1982)




\bibitem{Alvarez-Gaume:1984zst}
L.~Alvarez-Gaume, S.~Della Pietra and G.~W.~Moore,
``Anomalies and Odd Dimensions,''
Annals Phys. \textbf{163}, 288 (1985)

\bibitem{Witten:2015aba}
E.~Witten,
``Fermion Path Integrals And Topological Phases,''
Rev. Mod. Phys. \textbf{88}, no.3, 035001 (2016)
[arXiv:1508.04715 [cond-mat.mes-hall]].

\bibitem{Seiberg:1996bd}
N.~Seiberg,
``Five-dimensional SUSY field theories, nontrivial fixed points and string dynamics,''
Phys. Lett. B \textbf{388}, 753-760 (1996)
[arXiv:hep-th/9608111 [hep-th]].

\bibitem{Morrison:1996xf}
D.~R.~Morrison and N.~Seiberg,
``Extremal transitions and five-dimensional supersymmetric field theories,''
Nucl. Phys. B \textbf{483}, 229-247 (1997)
[arXiv:hep-th/9609070 [hep-th]].

\bibitem{Douglas:1996xp}
M.~R.~Douglas, S.~H.~Katz and C.~Vafa,
``Small instantons, Del Pezzo surfaces and type I-prime theory,''
Nucl. Phys. B \textbf{497}, 155-172 (1997)
[arXiv:hep-th/9609071 [hep-th]].

\bibitem{Katz:1996fh}
S.~H.~Katz, A.~Klemm and C.~Vafa,
``Geometric engineering of quantum field theories,''
Nucl. Phys. B \textbf{497} (1997), 173-195
[arXiv:hep-th/9609239 [hep-th]].

\bibitem{Intriligator:1997pq}
K.~A.~Intriligator, D.~R.~Morrison and N.~Seiberg,
``Five-dimensional supersymmetric gauge theories and degenerations of Calabi-Yau spaces,''
Nucl. Phys. B \textbf{497} (1997), 56-100
[arXiv:hep-th/9702198 [hep-th]].

\bibitem{Aharony:1997bh}
O.~Aharony, A.~Hanany and B.~Kol,
``Webs of (p,q) five-branes, five-dimensional field theories and grid diagrams,''
JHEP \textbf{01} (1998), 002
[arXiv:hep-th/9710116 [hep-th]].

\bibitem{Jefferson:2017ahm}
P.~Jefferson, H.~C.~Kim, C.~Vafa and G.~Zafrir,
``Towards Classification of 5d SCFTs: Single Gauge Node,''
[arXiv:1705.05836 [hep-th]].

\bibitem{Jefferson:2018irk}
P.~Jefferson, S.~Katz, H.~C.~Kim and C.~Vafa,
``On Geometric Classification of 5d SCFTs,''
JHEP \textbf{04}, 103 (2018)
[arXiv:1801.04036 [hep-th]].

\bibitem{Closset:2018bjz}
C.~Closset, M.~Del Zotto and V.~Saxena,
``Five-dimensional SCFTs and gauge theory phases: an M-theory/type IIA perspective,''
SciPost Phys. \textbf{6} (2019) no.5, 052
[arXiv:1812.10451 [hep-th]].


\bibitem{Lawrence:1997jr}
A.~E.~Lawrence and N.~Nekrasov,
``Instanton sums and five-dimensional gauge theories,''
Nucl. Phys. B \textbf{513} (1998), 239-265
[arXiv:hep-th/9706025 [hep-th]].

\bibitem{Candelas:1990rm}
P.~Candelas, X.~C.~De La Ossa, P.~S.~Green and L.~Parkes,
``A Pair of Calabi-Yau manifolds as an exactly soluble superconformal theory,''
Nucl. Phys. B \textbf{359}, 21-74 (1991)


\bibitem{Hori:2000kt}
K.~Hori and C.~Vafa,
``Mirror symmetry,''
[arXiv:hep-th/0002222 [hep-th]].




\bibitem{Witten:1997sc}
E.~Witten,
``Solutions of four-dimensional field theories via M theory,''
Nucl. Phys. B \textbf{500}, 3-42 (1997)
doi:10.1016/S0550-3213(97)00416-1
[arXiv:hep-th/9703166 [hep-th]].


\bibitem{Henningson:1997hy}
M.~Henningson and P.~Yi,
``Four-dimensional BPS spectra via M theory,''
Phys. Rev. D \textbf{57}, 1291-1298 (1998)
[arXiv:hep-th/9707251 [hep-th]].

\bibitem{Mikhailov:1997jv}
A.~Mikhailov,
``BPS states and minimal surfaces,''
Nucl. Phys. B \textbf{533}, 243-274 (1998)
[arXiv:hep-th/9708068 [hep-th]].

\bibitem{Hanany:1996ie}
A.~Hanany and E.~Witten,
``Type IIB superstrings, BPS monopoles, and three-dimensional gauge dynamics,''
Nucl. Phys. B \textbf{492}, 152-190 (1997)
[arXiv:hep-th/9611230 [hep-th]].

\end{thebibliography}
\end{document}